\documentclass[useAMS,usenatbib,twocolumn]{mn2e}

\usepackage{amsmath,amssymb}
\usepackage{graphicx}
\usepackage{natbib}

\newcommand{\spose}[1]{\hbox to 0pt{#1\hss}}

\newcommand{\seq}{\,{=}\,}

\mathchardef\star="313F

\newcommand{\aj}{AJ}         
\newcommand{\apj}{ApJ}       
\newcommand{\mnras}{MNRAS}   
\newcommand{\apss}{Ap\&SS}   
\newcommand{\nat}{Nat}       

\title[M-SSA]{Using Multichannel Singular Spectrum Analysis to Study
  Galaxy Dynamics}

\author[Weinberg \& Petersen]{Martin D. Weinberg\thanks{E-mail:
    weinberg@astro.umass.edu}$^{1}$, Michael~S.~Petersen$^{2}$\\
$^{1}$  Department of Astronomy, University of Massachusetts, Amherst
  MA 01003-9305, USA\\
$^2$Institute for Astronomy, University of Edinburgh, Royal Observatory, Blackford Hill, Edinburgh EH9 3HJ, UK }

\begin{document}

\label{firstpage}

\pagerange{\pageref{firstpage}--\pageref{lastpage}} \pubyear{2020}

\maketitle

\begin{abstract}
\(N\)-body simulations provide most of our insight into the structure
and evolution of galaxies, but our analyses of these are often
heuristic and from simple statistics. We propose a method that
discovers the dynamics in space and time together by finding the most
correlated temporal signals in multiple time series of basis function
expansion coefficients and any other data fields of interest.  The
method extracts the dominant trends in the spatial variation of the
gravitational field along with any additional data fields
\emph{through} time.  The mathematics of this method is known as
\emph{multichannel singular spectrum analysis} (M-SSA).  In essence,
M-SSA is a principal component analysis of the covariance of time
series replicates, each lagged successively by some interval. The
dominant principal component represents the trend that contains the
largest fraction of the correlated signal.  The next principal
component is orthogonal to the first and contains the next largest
fraction, and so on. Using a suite of previously analysed simulations,
we find that M-SSA describes bar formation and evolution, including
mode coupling and pattern-speed decay. We also analyse a new
simulation tailored to study vertical oscillations of the bar using
kinematic data. Additionally, and to our surprise, M-SSA uncovered
some new dynamics in previously analysed simulations, underscoring the
power of this new approach.
\end{abstract}

\begin{keywords}
  methods: numerical --- galaxies: simulation, structure
\end{keywords}

\section{Introduction}
\label{sec:intro}

For the last fifty years, most of our insight about the dynamics of
evolving galaxies has come from \(N\)-body simulations.  Indeed, one
of the earliest \(N\)-body simulations of galaxies immediately
revealed the bar instability \citep{Hohl:1971}.  We design these
numerical experiments to explore scenarios such as stability or
interaction. Then, we observe movies of the various projections of
body density or smoothed distributions in time to identify salient
dynamical features.  Scientists are trained to detect patterns and the
human brain-eye combination is adept at finding complex relationships.
This leads to qualitative interpretation, quantified perhaps by simple
statistics.  However, humans are easily distracted by the `big thing'
in the frame. Owing to this, much of the dynamical information in our
simulations is untapped. Can we extract unseen or low-amplitude but
significant dynamics from simulations?  Can we find algorithms that
summarise the dynamical degrees of freedom even if we do not know them
to start? These two questions motivated this paper.

To study this, one must represent the dynamics in coordinates that
focus on the mechanism. Action-angle expansions do this very well and
are useful for systems which do not evolve with time
\citep{Binney.Tremaine:2008}.  However, we know that galaxies
\emph{do} evolve in time, and this time-dependence influences the
evolution itself.  The nature of galactic dynamics requires analysis
that can accurately treat multiple time scales.  For example,
interesting evolution takes place on time scales longer than the
characteristic period, largely driven by resonant exchange and phase
mixing.  To do this efficiently, one may follow the gravitational
field in evolving systems using basis-function expansions (BFE).  BFE
represents the potential and density of near-equilibrium galaxies in
basis functions: pairs of biorthogonal functions that satisfy the
Poisson equation and are mutually orthogonal \citep{Clutton-Brock:72,
  Clutton-Brock:73, Fridman.Polyachenko:84b, Hernquist.Ostriker:92,
  Weinberg:99, Lilley.etal:2018a, Lilley.etal:2018b}.  As described in
\citet{Weinberg:99}, we select our basis functions to best represent
the underlying galaxy density profile.  In effect, this windows the
length scales to particular ones of interest.  Additionally useful,
one can take any set of biorthogonal functions and create a new
biorthogonal set of functions by a vector-space rotation. In
particular, one may select the rotation by diagonalising the variance
matrix of the expansion coefficients given the particle distribution;
this generates a new set of orthogonal functions that looks like the
particle distribution \citep{Petersen.etal:2019a}.  This new
data-motivated basis is sometimes called \emph{empirical orthogonal
  functions} (EOFs) after \cite{Lorenz:56}.

In short, a biorthogonal basis can be made to represent the individual
degrees of freedom in a simulation's spatial structure at one point in
time.  The full dynamical simulation is a time series of such
representations.  Although we require many time slices for a accurate
solution to the equations of motion, interesting dynamics correlate
these fields on large, natural time scales.  When represented by a
BFE, the information become time series of coefficients.  Can we use
the information in these time series to discover the dynamics by
analysis, in much the same way that we construct EOFs?

The answer is unambiguously {\sl yes}.  The simplest approach might be
an application of Fourier analysis of these time series to identify
the characteristic time scales followed by filtering to reconstruct
series that capture the power.  However firstly, Fourier analysis is
challenging for time series with periods that are not very short
compared to the entire time segment.  Secondly, Fourier analysis works
best as a filter for known signals of well-defined and
time-independent period.  That is not the case for evolving galaxies.
To address both of these concerns we adopt a more general data-driven
method called \emph{Singular Spectrum Analysis} (SSA).  SSA is a
non-parametric analysis tool for for time series analysis.  It was
introduced into the geophysical literature by
\citet{Ghil.Vautard:1991}.  Subsequently, has been used for a wide
variety of primarily Earth-science problems where finite data sets
that may be censored are prevalent (see \citealt{Ghil.etal:2002} for a
comprehensive review and the monograph
\citealt{Golyandina.Zhigljavsky:2013} for mathematical detail).  It
makes no strong prior assumptions about the spectrum.  An oscillation
does not have to be periodic to be accurately represented.  For
example, an oscillator with a decreasing period is not spread over
multiple frequency channels as in Fourier analysis but represented as
a single component in SSA.  We will show an example of this in
representing a barred galaxy with a decreasing pattern speed in
section \ref{sec:barfid}.  There is a natural multivariate extension
(called \emph{multichannel} or \emph{multivariate} in the SSA
literature; M-SSA) that we will use below.

In many geophysical applications, the time series is a direct
observable.  The main extension here is to use the spatial summary
provided by BFE coefficients as the multichannel
data.  The subsequent analysis is then a combined spatio-temporal
filter and reconstruction of the dynamics.  This is exactly what we
need and want for \(N\)-body dynamical analysis.  Specifically: each
expansion coefficient represents a feature in the gravitational field
at some specific scale.  Thus, the analysis of multiple time series of
BFE coefficients using SSA automatically discovers the key
correlations in space \emph{and} time without supervision (in the
machine-learning sense).

The plan of this paper is as follows.  We will begin with quick
motivation and review of BFEs (section \ref{sec:BFE}).  These methods
have been described in many places (op. cit.); here, we will emphasise
the adaptive non-parametric aspects of this technique.  This will be
followed by a brief introduction to SSA and the multichannel
generalisation, M-SSA (section \ref{sec:ssaintro}). This is only meant
to present the basic idea; we refer the reader to a recent monograph
\citep{Golyandina.Zhigljavsky:2013} for a complete discussion.
Finally, section \ref{sec:examples} applies the M-SSA analysis to the
BFE time series for several standard simulations that form stellar
bars to illustrate the use and power of this method.

Section \ref{sec:barfid} applies M-SSA to the simulation studied by
\citet{Petersen.etal:2019b}. This paper identifies specific time
intervals or \emph{epochs} where the dynamics is dominated by a
particular mechanism.  We find that M-SSA identifies the same epochs
with no prior information.  This simulation also reveals a novel
coupling between the quadrupole excitation caused by the stellar bar
and the dipole (\(m=1\)) seiche mode.  By using the combination of
\(m=1\) and \(m=2\) BFE coefficients in section \ref{sec:m12}, we
recover these dynamics.  In addition, M-SSA reveals a previously
undetected feature: the excitation of a very slow retrograde \(m=1\)
similar to that predicted in \citet{Weinberg:94c}.  This feature is
very hard to see by eye in any movie, but it clearly detected with
considerable power by M-SSA.

One limitation of the standard BFE approach is its limitation to
observables only in the density-potential field.  However, using
M-SSA, we can adjoin time series that describes any summary
information we desire.  As an example, in section \ref{sec:kine}, we
add time series of kinematic data for a stellar disc from the same
simulations in one of two ways: a spatial Fourier analysis in radial
rings an a Fourier-Laguerre analysis over a finite disc.  We show that
the latter method accurately couples the expected kinematic field of
trapped bar orbits.  Section \ref{sec:buckle} applies the combined
BFE--kinematic time series to analyse bar \emph{buckling} for a
simulation with thickened stellar disc.  M-SSA efficiently identifies
distinct vertical evolution and demonstrates that the dynamics are
tied to the bar.

The examples in section \ref{sec:examples} were chosen because we have
already studied them in detail but were complex enough that they might
illustrate real galaxy dynamics applications with BFE and M-SSA.  We
feel that they have done that, as well as yield unexpected new
dynamics and insight. This paper just begins to scratch the surface of
what might be possible with this method.  We end with a summary of
results and discussion of future directions in both application
theoretical development in section \ref{sec:discuss}.

\section{Quick introduction to basis-function expansions (BFE)}
\label{sec:BFE}

A BFE computes the gravitational potential by projecting particles
onto a set of biorthogonal basis functions that satisfy the Poisson
equation.  Then, the force at the position of each particle is
evaluated from the basis-function approximation to the field at the
particle position.  Fundamentally, this approach relies on the
mathematical properties of the Sturm-Liouville equation (SLE) of which
the Poisson equation is a special case. The SLE describes many
physical systems, and may be written as:
\begin{equation}
\frac{d}{dx}\left[p(x)\frac{d\Phi(x)}{dx}\right] -
q(x)\Phi(x) \seq \lambda w(x) \Phi(x)
\label{eq:sle}
\end{equation}
where \(\lambda\) is a constant, and \(w(x)>0\) is a weighting
function. The eigenfunctions \(\phi_j\) of the SLE form a complete
basis set with eigenfunctions \(\lambda_j\). The BFE potential solver
is built using properties of eigenfunctions and eigenvalues of the
SLE.

For an example of this for an approximately spherical system, begin by
separating the Poisson equation in spherical coordinates.  The angular
degrees of freedom in this separation are satisfied by spherical
harmonics, \(Y_{lm}\) and the remaining radial equation has the form
of equation (\ref{eq:sle}).  The weighting function \(w(r)\) in
equation (\ref{eq:sle}) may be selected to provide an equilibrium
solution of the Poisson equation.  Then, the unperturbed potential
would be represented by a single basis function!  To do this, we choose a
weighting function \(w(r)\propto \rho_0(r)\Phi_0(r)\) where
\(\rho_0(r),\Phi_0(r)\) is the unperturbed model. Substituting
\(\Phi(r)=\Phi_0(r)u(r)\) in the Poisson equation with \(x=r\), the
coefficients in equation (\ref{eq:sle}) become:
\begin{eqnarray} p(r) &=& r^2\Phi_0^2(r) \\ q(r) &=&
  \left[l(l+1)\Phi_0(r) - \nabla_r^2\Phi_0(r)r^2\right]\Phi_0(r)
  \\ w(r) &=& -4\pi G r^2\Phi_0(r)\rho_0(r).
\end{eqnarray}
By changing the weighting function \(w(r)\) and applying physical
boundary conditions\footnote{One may also precondition the
  eigenfunction \(\Phi(x) = f(x)\phi(x)\) for some well-behaved
  function \(f(x)\) instead of changing the weighting function.}, we
may derive an infinity of radial bases in both finite and infinite
domains.  Each eigenfunction \(u^{j}_{lm}(r)\) for a particular \(l,
m\) then corresponds to a solution of equation~(\ref{eq:sle}). The
lowest-order eigenfunction with \(\lambda=1\) is \(u_{00}^{0}(r)=1\)
by construction.  Each successive $u_{lm}^{j}(r)$ with $j>0$ has an
additional radial node. The resulting series of eigenfunctions
\(u^{j}_{lm}(r)\) correspond to the potentials
\(\phi^{j}_{lm}\propto\Phi_0(r)u^{j}_{lm}(r)\). Each term in halo
potential is then given by \(\Phi_{lm}^j \seq
\phi_{lm}^j(r)Y_{lm}(\theta,\phi)\).  The radial boundary conditions
are straightforward to apply at the origin, at infinity, or wherever
you like. Then, a near-spherically-symmetric system, such as a
dark-matter halo, can be expanded into a relatively small number of
spherical harmonics and appropriately chosen radial functions.

The disc is more complicated.  Although one can construct a disc basis
from the eigenfunctions of the Laplacian as in the spherical case
\citep[e.g.][]{Earn:96}, the boundary conditions in cylindrical
coordinates make the basis hard to implement. Our method uses singular
value decomposition (SVD) of a high-order (\(l\le36\)) spherical basis
to define a rotation in function space to best represent a target disc
density.  Specifically, each density element \(\rho(R, z)\,d^3x\)
contributes
\begin{equation}
\frac{1}{4\pi G}\phi_{lm}^j(r)Y_{lm}(\theta,\phi)\rho(R, z)d^3 x
\end{equation}                                      
to the expansion coefficient \(a_{lm}^j\), or
\begin{eqnarray}
  a_{lm}^j &=& \frac{1}{4\pi
    G}\int \phi_{lm}^j(r)Y_{lm}(\theta,\phi)\rho(R, z)d^3x \nonumber \\
   \hat{a}_{lm}^j &=& \lim_{N\rightarrow\infty}\frac{1}{4\pi
    G}\sum_{i=1}^N m_i \phi_{lm}^j(r_i)Y_{lm}(\theta_i,\phi_i)
  \label{eq:expcof}
\end{eqnarray}
where \(R, z\) are the radial and vertical cylindrical coordinates.
The second equation shows the approximation for \(N\) particles where
\(\sum_i m_i = \int \rho(R, z)d^3x\).  We represent quantities
estimated from the particles with \(\hat{\cdot}\).  The potential and
density are then estimated as follows:
\begin{eqnarray}
  \hat{\rho}(\mathbf{x}) &=& \sum_{jlm} \hat{a}_{lm}^j \rho_j^{lm}(r)Y_{lm}(\theta,
  \phi), \label{eq:expden} \\
  \hat{\Phi}(\mathbf{x}) &=& \sum_{jlm} \hat{a}_{lm}^j \phi_j^{lm}(r)Y_{lm}(\theta,
  \phi). \label{eq:exppot}
\end{eqnarray}

The covariance of the coefficient given the density \(\rho(R, z)\),
\(\mbox{cov}(\mathbf{a})\), is constructed similarly.  This covariance
is \emph{not} the classic variance about the mean but rather the
variance about zero.  In this way, the covariance matrix describes
which terms \(a_{lm}^j\) contribute the most to the gravitational
field overall; converted to physical units, the \(|a_{lm}^j|^2\)
represent gravitational energy. The diagonalisation by SVD of
\(\mbox{cov}(\mathbf{a})\) provides a new basis that is uncorrelated
by the target density.  Because \(\mbox{cov}(\mathbf{a})\) is
symmetric and positive definite, all eigenvalues will be positive.
The term with the largest eigenvalue describes the majority of the
correlated contribution, and so on for the second largest eigenvalue,
etc.  The singular matrices from the decomposition (now mutual
transposes owing to symmetry) describe a rotation of the original
basis into the uncorrelated basis.

The new basis functions optimally approximate the true distribution
from the spherical-harmonic expansion in the original basis in the
sense that the largest amount of variance or \emph{power} in the
gravitational field is contained in the smallest number of terms.  We
might call this optimal in the \emph{least-squares sense} since the
SVD solution of the least-squares problem provides the same optimal
properties.  More importantly, the transformation and the Poisson
equation are linear, and therefore the new eigenfunctions are also
biorthogonal.  The new coefficient vector is related to the original
coefficient vector by an orthogonal transformation.  Because we are
free to break up the spherical basis into meridional subspaces by
azimuthal order, the resulting two-dimensional eigenfunctions in \(r\)
and \(\theta\) are equivalent to a decomposition in cylindrical
coordinates \(r,~z,\) and \(\theta\).  For the applications in this
paper, we condition the initial disc basis functions on an analytic
disc density such that the lowest-order potential-density pair matches
the initial analytic mass distribution. This choice also acts to
reduce small-scale discreteness noise as compared to conditioning the
basis function on the realised positions of the particles
\citep{Weinberg:98a}.

We are free to represent the potential and density of a galaxy as a
superposition of several families of basis functions. This allows us
to decompose the galaxy into components of different geometry and
symmetry. For an initially axisymmetric example, azimuthal harmonics
\(m\), where \(m\seq0\) is the monopole, \(m\seq1\) is the dipole,
\(m\seq2\) is the quadrupole, and so on, will efficiently summarise
the degree and nature of the asymmetries. The sine and cosine terms of
each azimuthal order give the phase angle of the harmonic that can be
used to calculate the pattern speed.  For discs, each term in the
expansion at a particular azimuthal order \(m\) represents both the
radial and vertical structure simultaneously; that is, each basis
function is a two-dimensional meridional plane multiplied by
\(e^{im\phi}\).  The symmetry of the input basis and the covariance
matrix further demands that the SVD produce vertically symmetric or
antisymmetric functions about the \(z=0\) plane.

The BFE approach trades off precision and degrees-of-freedom with
adaptability. The truncated series of basis functions intentionally
limits the possible degrees of freedom in the gravitational field in
order to provide a low-noise bandwidth-limited representation of the
gravitational field.  A simulation performed with any Poisson solver
may be summarised using a BFE: a basis-function representation
provides an information-rich summary of the gravitational field and
provides insight into the overall evolution.  This method allows for
the decomposition of different components into dynamically-relevant
subcomponents for which the gravitational field can be calculated
separately. 

For the purposes of this paper, the coefficients in equation
(\ref{eq:expcof}) are time series that represent the spatial density
and gravitational potential fields of a dynamical simulation
\citep[presented in][]{Petersen.etal:2019a}.  Note
that the mathematics that allows us to find a high-dimensional
rotation in the vector space of coefficients to obtain a cylindrical
expansion from a spherical expansion is a form of \emph{unsupervised
  learning}.  That is, we allow the dynamical simulation to serve as
the data that best determines a best basis to represent itself.

The SSA method, described in the next section, uses the same general
idea as the spatial EOFs.  Specifically, we can generate EOFs from
biorthogonal functions that represent most of the correlated
gravitational energy in a small number of degrees of freedom.  We can
do the same with series of samples in the time domain: by analysing
the variance of the time series with itself or with others at various
time intervals, we can find functions that describe most of the
correlation.  The main contribution of this paper is putting the EOFs
in both space and time together in a single analysis.  That is, the
SSA algorithm allows the same mathematical principles to determine the
best representation of the time-series coefficients from equation
(\ref{eq:expcof}).  In this way, our analyses described in section
\ref{sec:examples} will be a combined spatio-temporal representation
of the key dynamics in our simulation.

\section{SSA algorithms and methodology}
\label{sec:ssaintro}

SSA analysis separates the observed time series into the sum of
interpretable components with no a priori information about the time
series structure. We begin with a statement of the underlying
algorithm for a single time series following the development in
\citet{Golyandina.Zhigljavsky:2013}.  In other words, we consider one
particular coefficient \(a_j(t)\) from equation (\ref{eq:expcof}) at a
particular time step.  Let us simply denote the coefficient at time
step \(k\) as \(a_{j,k} = a_j(t_o+hk)\) where \(h\) is the time-step
interval.

\subsection{The SSA Algorithm}

Since we are considering a single coefficient \(a_j(t)\), we will drop
the coefficient index \(j\) for now.  Denote the real-valued time
series of coefficients \(a_{j,k}\) of length \(N\) as
\(\mathbf{a}_N=(a_1,\ldots,a_{N})\) of length \(N\).  The SSA
algorithm (1) decomposes the temporal cross-correlation matrix by an
eigenfunction analysis into uncorrelated components and then (2)
reconstructs relevant parts of the time series.  Each of the
subsections below gives the mathematical expressions needed to
accomplish this together with some explanation of each step.

\subsubsection{Decomposition}
\label{subsect:decomp}

\paragraph{\textbf{Embedding}.}
We \emph{embed} the original time series into a sequence of lagged
vectors of size \(L\) by forming \(K=N-L+1\) {\em lagged vectors}
\[
A_i=(a_{i},\ldots,a_{i+L-1})^\top, \quad i=1\ldots,K.
\]
This new data matrix of lagged vectors is often called the
\emph{trajectory matrix} in the SSA literature.  The \(L\times K\)
\emph{trajectory matrix} of the series \(A_N\) is
\begin{eqnarray}
  \label{eq:traj_m}
  \mathbf{T} &=& [A_1:\ldots:A_K]=(T_{ij})_{i,j=1}^{L,K} \nonumber \\
  &=&
  \left(
  \begin{array}{lllll}
    a_1&a_2&a_3&\ldots&a_{K}\cr
    a_2&a_3&a_4&\ldots&a_{K+1}\cr
    a_3&a_4&a_5&\ldots&a_{K+2}\cr
    \vdots&\vdots&\vdots&\ddots&\vdots\cr
    a_{L}&a_{L+1}&a_{L+2}&\ldots&a_{N}\cr
  \end{array}
  \right).
\end{eqnarray}
There are two important properties of the trajectory matrix: the rows
and columns of \(\mathbf{T}\) are subseries of the original series,
and \(\mathbf{T}\) has equal elements on anti-diagonals\footnote{This is
called a Hankel matrix in linear algebra texts.}.   For the practitioner,
these constant skew-diagonals must be maintained in order to
reconstruct features of the original time series.

From the trajectory matrix, we can form the \emph{lag-covariance}
matrix:
\begin{equation}
  \label{eq:lagcovar}
  \mathbf{C} = \frac{1}{K} \mathbf{T}^\top\cdot\mathbf{T}.
\end{equation}
This matrix will be analysed by SVD to find the representation in the
lagged time space that best represents the correlations between the
input data at different time lags\footnote{An alternative
  decomposition that is commonly used in the SSA literature is based
  on the eigenvectors of the Toeplitz matrix \(\mathbf{C}\) whose
  entries are
  \[
  c_{ij}=\frac{1}{N-|i-j|} \sum_{n=1}^{N-|i-j|}a_n a_{n+|i-j|}, \quad
  1\leq i,j\leq L.
  \]
  The Toeplitz formulation reduces approximately to the covariance
  form for stationary time series with zero mean.  and can also be
  decomposed in \(\mathcal{O}(N^2)\) time in contrast with the SVD
  which requires \(\mathcal{O}(N^3)\) time.  However, \(N\)-body
  simulations are evolving systems and not time stationary.}.  If the
signal were periodic, the \emph{window length} \(L\) roughly describes
the maximum period that could be identified by cross correlation.

As an example, consider the trajectory matrix for a pure sinusoidal
time series, \(\sin t\), that is much longer than the
period\footnote{We will continue to explore this example in
  Section~\protect{\ref{sec:sinusoid}}.}.  The computation of the lag
covariance matrix in equation (\ref{eq:lagcovar}) will correlate every
row with every column.  At lags \(0, 2\pi, 4\pi, \ldots\), the
correlation is purely constructive and positive.  At lags \(\pi, 3\pi,
5\pi, \ldots\), the correlation is purely constructive and negative.
At lags \(\pi/2, 3\pi/2, 5\pi/2, \ldots\), the correlation is purely
destructive, and so on.  Altogether, the cross correlation of the
trajectory matrix will have elements \(c_{jk} \rightarrow
\cos(h(j-k))/2\) in the limit of large \(K\).

\paragraph{\textbf{Decomposition}.}

We analyse the lag-covariance matrix using the SVD.  From the form of
equation (\ref{eq:lagcovar}), we observe that \(\mathbf{C}\) is real,
symmetric and positive definite, so the SVD yields a decomposition of
the form: \(\mathbf{C} =
\mathbf{U}\cdot\mathbf{\Lambda}\cdot\mathbf{V}^\top\) where
\(\mathbf{\Lambda}\) is diagonal. The symmetry properties imply that
the left- and right-singular vectors are the same, or
\(\mathbf{E}\equiv\mathbf{U}=\mathbf{V}\).  We may then write
\begin{equation}
  \mathbf{\Lambda} = \mathbf{E}^\top\cdot\mathbf{C}\cdot\mathbf{E}.
  \label{eq:diag}
\end{equation}
The matrix \(\mathbf{\Lambda}\) is a diagonal matrix of eigenvalues,
\(\lambda_k\) and the columns of \(\mathbf{E}\) are the eigenvectors,
\(\mathbf{E}^k\). For problems considered here, the rank of the
covariance matrix is usually under several thousand so the numerical
work in performing the SVD is manageable using the divide-and-conquer
algorithm \citep{Gu.Eisenstat:2006} without special-purpose hardware.

After performing the SVD, we have a decomposition into eigenvalues and
eigenvectors.  The pair \((\sqrt{\lambda_k}, \mathbf{E}^k)\) is called
the \(k\)th \emph{eigenpair}. Let us assume that the eigenpairs are
sorted in order of decreasing value of \(\lambda_k>0\), which is
traditional for SVD. As before, we may write this decomposition in
\emph{elementary matrix} form as
\begin{equation}
\label{eq:elem_matr}
\mathbf{C} = \sum_k \lambda_k \mathbf{E}^k \mathbf{E}^{k\top}
           = \sum_k \lambda_k \mathbf{E}^k \otimes \mathbf{E}^{k}   
           = \sum_k \mathbf{C}_k
\end{equation}
where \(\mathbf{a}\otimes\mathbf{b}\) denotes the outer or Kronecker
product of the vectors \(\mathbf{a}\) and \(\mathbf{b}\) and
\(\mathbf{C}_k \equiv \lambda_k \mathbf{E}^k\otimes\mathbf{E}^k\).
Clearly, the \(\mathbf{C}_k\) have dimension \(K\times K\).

It is also useful to define the normalised cumulative sum
of the sorted eigenvalues,
\begin{equation}
\mathcal{S}_{K^\prime} \equiv \frac{\sum_{k=1}^{K^\prime}
  \lambda_k}{\sum_{k=1}^K \lambda_k},
\label{eq:eigensum}
\end{equation}
where \(K^\prime<K\), to quantify the relative importance of
eigenvectors.  

Let us again consider our simple sinusoidal example. The sum in
equation (\ref{eq:elem_matr}) would reduce our lag covariance matrix
\(\mathbf{C}\) to a single elementary matrix.

\subsubsection{Reconstruction}

\paragraph{\textbf{The principal components}.}
Using the symmetry of \(\mathbf{T}\), equations (\ref{eq:lagcovar})
and (\ref{eq:diag}) imply that
\begin{equation}
  \mathbf{\Lambda} =
  \frac{1}{K}\left(\mathbf{E}^\top\cdot\mathbf{T}\right)
  \cdot\left(\mathbf{E}^\top\cdot\mathbf{T}\right)^\top.
\end{equation}
Since \(\mathbf{\Lambda}\) is diagonal, the trajectory matrix in the
basis of the eigenvectors \(\mathbf{E}\) must be spanned by the set of
orthogonal vectors: \(\mathbf{P} = \mathbf{E}^\top\cdot
\mathbf{T}\). The vectors, which are columns of \(\mathbf{P}\), are
known as the \emph{principal components} (PCs), following the
terminology of standard Principal Component Analysis (PCA).  Each
eigenpair yields a single PC.  The matrix \(\mathbf{E}\) transforms
the lagged time series to a unlagged representation.  Each PC
represents some part of the signal at zero lag from the original time
series that is uncorrelated (orthogonal) to any other PC.  The value
of the PC for eigenpair \(k\) at time step \(j\) is
\begin{equation}
  \label{eq:pc1d}
  P^k_j = \sum_{l=1}^L E^k_l T_{lj}  = \sum_{l=1}^L E^k_l a_{j+l-1}.
\end{equation}

Again considering our simple sinusoidal example, the PCs will be
dominated by one pair of PCs and the rest of the components will have
zero eigenvalues.  You would be correct in complaining that this is an
overly elaborate way of performing a periodogram.  The power in the
method is that PCs need not be sinusoidal and SSA will still decompose
the signal as a set of uncorrelated, orthogonal parts.

\paragraph{\textbf{The reconstructed components}.}
\label{sec:reconstructed}

This singular value decomposition is the \emph{best} separation of
correlated temporally varying signal in a peculiar vector space which
removes the time lag.  Now, we want to relate the contribution of each
separated component to the original time series for physical
interpretation. To do this, we use again the unitary transformation
defined by our eigenvectors, \(\mathbf{E}\), to reconstruct our
trajectory matrices that describe the original time series.  Recall
that each unique coefficient in the trajectory matrix appears many
times along the anti-diagonals (the \emph{Hankel} property).  However,
this property is not guaranteed by the transformation.  Therefore, to
recover a consistent approximation of the contribution of a component
to the original time series, we diagonally average the result,
imposing the Hankel property of the input trajectory matrix.

Specifically, the transformed PCs corresponding to the eigenpair \(k\)
are: \(\tilde{\mathbf{T}}^k = \mathbf{P}^\cdot\cdot\mathbf{E}^k\). We
will use \(\tilde{\cdot}\) to denote quantities reconstructed from the
PCs.  Making the \emph{anti-diagonal average} to get the reconstructed
coefficients, we have:
\begin{equation}\small
  \tilde{a}^k_j =
  \begin{cases} \displaystyle
    \frac{1}{j} \sum_{n=1}^{j} P^k_{n-j+1} E^k_n & \mbox{if}\ 1\le j < L-1, \\
    \displaystyle
    \frac{1}{L} \sum_{n=1}^{L} P^k_{n-j+1} E^k_n & \mbox{if}\ L\le j \le N - L + 1 \, \\
    \displaystyle
    \frac{1}{N-j+1} \sum_{n=N-L+1}^{N} P^k_{n-j+1} E^k_l & \mbox{if}\ N-L+2\le j \le N. \\
 \end{cases}
\end{equation}

\begin{figure}
    \centering
    \includegraphics[width=0.5\textwidth]{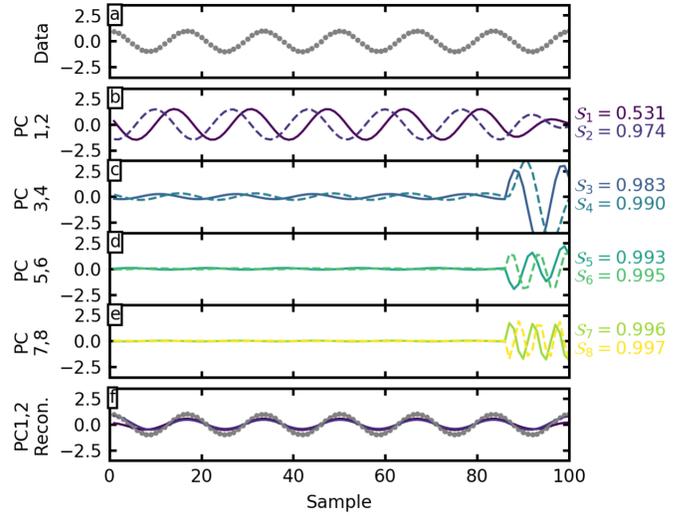}
    \caption{The cosine SSA example with \(L=15\). Panel a shows the
      cosine data. Panels b-e show the first eight PCs, grouped into
      four pairs. Alongside each panel we list the normalised
      cumulative sum of the eigenvalues.  Panel f shows the
      reconstruction of the first two PCs (coloured curves) and the
      sum of the first two PCs (grey) overlaid on the data.
      \label{fig:cosineL15}}
\end{figure}

\begin{figure}
    \centering
    \includegraphics[width=0.5\textwidth]{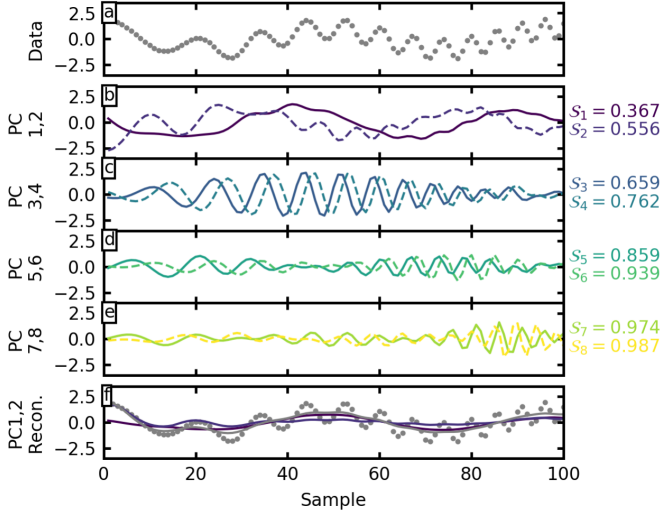}
    \caption{Same as Figure~\protect{\ref{fig:cosineL15}}, but for the
      chirp SSA example.
      \label{fig:chirpL15}}
\end{figure}

\begin{figure}
  \centering
  \includegraphics[width=0.5\textwidth]{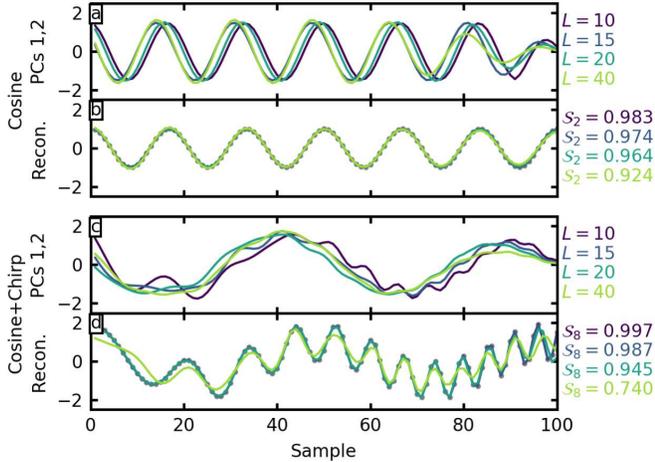}
  \caption{The effect of window sizes in the cosine example (panels a
    and b) and chirp (panels c and d). Panels a and c show the first
    two PCs for four different window lengths \(L\). Panels b and d
    show the reconstruction from the first two PCs. The input data is
    shown in points.  The normalised cumulative sum of the first two
pp    eigenfunctions is listed alongside.
    \label{fig:windows}}
\end{figure}

\begin{figure}
  \centering
  \includegraphics[width=0.4\textwidth]{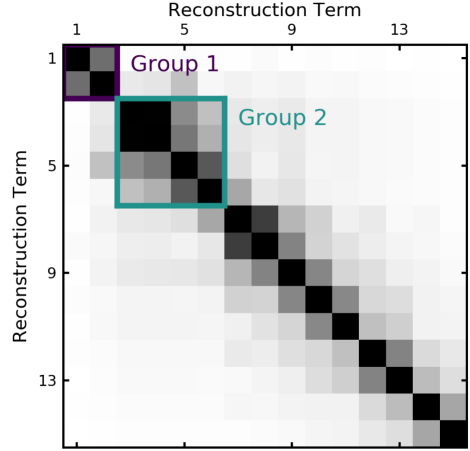}
  \caption{The distance matrix for the \(L=15\) cosine + chirp example
    with low (high) distance represented as black (white).  The top
    2-by-2 block is the cosine- and sine-like PCs from the pure
    sinusoid (group 1; outlined in dark blue).  The next 4-by-4 block
    contains the PCs that represent the chirp (group 2; outlined in
    green).  These first 6 PCs contain 94\% of the total variance.
    The k-means algorithm finds three groups, breaking the second
    4-by-4 block into two 2-by-2 blocks.  All PCs beyond the first 8
    are insignificant; we show them here for completeness.
    \label{fig:wcorr}}
\end{figure}

\subsection{Separability}

\label{sec:separ}

While the eigenpairs represent the \emph{best} decomposition of the
lag covariance matrix in the linear-least-squares sense, the eigenpair
components are not guaranteed to represent unique dynamical features.
In many cases, a single physical mechanism is spread into multiple
components.  We must \emph{group} like components together to best
elicit the mechanism.  We will see an example of this in Section
\ref{sec:barfid} when apply this method to our fiducial barred-galaxy
simulation.

Our grouping procedure has two parts. First, we truncate the component
series by only keeping eigenpairs that contribute significantly.
Choosing how many eigenpairs or PCs to keep is a bit ambiguous
especially for noisy time series.  To be explicit, we define the
maximum index \(d\): \(d=\max\{k:\ \lambda_k > \epsilon\}\), where
\(\epsilon\) is some threshold for the eigenvector to be in the null
space.  The run of \(\log\lambda_k\) often has rapid drop off
following by slowly decreasing plateau.  A common heuristic truncates
at the beginning of the plateau.  This corresponds to the point of
maximum curvature (often called the \emph{shoulder point}) in the
normalised cumulative sum \(\mathcal{S}_{K^\prime}\)
(eq. \ref{eq:eigensum}).  This shoulder often corresponds to a
cumulative normalised variance \(\mathcal{S}_{K^\prime}>0.8\) and
commonly closer to \(\mathcal{S}_{K^\prime} \approx 0.95\).  We have
found it useful to characterise the truncation \(\epsilon\) by the
cumulative fraction of total variance at the truncation index.

Secondly, from the decomposition in equation (\ref{eq:elem_matr}), the
grouping procedure partitions the set of indices \(\{1,\ldots,d\}\)
into \(m\) disjoint subsets \(I_1,\ldots,I_m\).  Define
\(\mathbf{C}_I=\sum_{k\in I} \mathbf{C_k}\).  Equation
(\ref{eq:elem_matr}) leads to the decomposition
\begin{equation}
\label{eq:mexp_g}
\mathbf{C}=\mathbf{C}_{I_1}+\ldots+\mathbf{C}_{I_m}.
\end{equation}
The procedure of choosing the sets \(I_1,\ldots,I_m\) is called
\emph{eigenpair grouping}.  We will describe strategies for grouping
below.  If \(m=d\) and \(I_k=\{k\}\), \(k=1,\ldots,d\), then the
corresponding grouping is called \emph{elementary}.  The choice of
several leading eigenpairs corresponds to the approximation of the
time series in view of the well-known optimal property of the SVD.
Grouping for our simple sinusoidal example, of course, is trivial: one
pair of elementary matrices that contains all of the information about
the signal.

The goal of grouping into sets \(\{I_k\}\) is the separation of the
time series into distinct dynamical components.  Distinct time series
components can be identified based on their similar temporal
properties.  For example, if the underlying dynamical signals are
periodic, then the PCs will reflect that by producing sine- and
cosine-like pairs with distinct frequencies.  Thus, graphs of PCs or
discrete Fourier transforms can help identify like components.  For a
real-world application, a quick visual examination of the \(d\) PCs
will suggest a group partition.

Very helpful information for separation is contained in the so-called
\(\mathbf{w}\)-correlation matrix. This is the matrix consisting of
weighted correlations between the reconstructed time series
components. The weights reflects the number of entries of the time
series terms into its trajectory matrix.  Well separated components
have small correlation whereas badly separated components have large
correlation. Therefore, looking at the \(\mathbf{w}\)-correlation
matrix one can find groups of correlated elementary reconstructed
series and use this information to construct groups \(\mathbf{C_k}\).
The goal is to find groups such that uncorrelated reconstructed
trajectories belong to different groups.

Mathematically, the elements \(\mathbf{w}\)-correlation matrix are
related to the Euclidean-like matrix norm (also known as the
\emph{Frobenius} norm), defined as
\begin{equation}
||\mathbf{M}||_F = \sqrt{\sum_{i=1}^n \sum_{j=1}^m a_{ij}^2}
\label{eq:frobenius}
\end{equation}
for an \(n\times m\) matrix \(\mathbf{M}\) \citep{Golub.Loan:1996}.
Specifically, let \(\tilde{\mathbf{T}}^k\) be the reconstructed
trajectory matrix defined in section \ref{sec:reconstructed}.  Define
the normed trajectory matrix \(\mathcal{T}^k =
\tilde{\mathbf{T}}^k/||\tilde{\mathbf{T}}^k||_F\).  Then, the elements
of the \(\mathbf{w}\)-correlation matrix are
\begin{equation}
  w^{1/2}_{\mu,\nu} = 1 - \frac{1}{2}||\mathcal{T}^\mu -
  \mathcal{T}^\nu||_F.
  \label{eq:wcorr}
\end{equation}
For insight, note that the same construction defines a metric distance
\begin{equation}
  d_{\mu,\nu} = 1 - w^{1/2}_{\mu,\nu} = \frac{1}{2}||\mathcal{T}^\mu -
  \mathcal{T}^\nu||_F
  \label{eq:dist}
\end{equation}
between the two trajectory matrices.  So the
\(\mathbf{w}\)-correlation matrix describes the distance between
reconstructed trajectory matrices.  Thus, \(w_{\mu,\nu}=0\) if the
reconstructed series for PCs with indices \(\mu\) and \(\nu\) are
orthogonal and \(w_{\mu,\nu}=1\) if they are linearly dependent.  The
elements \(d_{\mu,\nu}\) are a matrix which is the complement of the
\(\mathbf{w}\)-correlation matrix which we will call the
\emph{distance} matrix.

From our experience, grouping is the trickiest and most sensitive part
of the method; undergrouping will spread a single process into
multiple features.  For example, consider time series of coefficients
from the basis EOFs from section \ref{sec:BFE} for some simulation.  A
single correlated dynamical signal may appear as several signals with
very similar but slightly different principal components.  For
examples considered in section \ref{sec:examples}, we have grouped by
examining the distance matrices to see which elements form obvious
blocks.  Often these blocks are obvious, and we provide an example in
section \ref{sec:sinusoid}.  At the very least, for a strong nearly
periodic perturbation such as a galaxy bar, nearly all significant
eigenpairs come in phase pairs, corresponding to the cosine- and
sine-like components of the oscillation.  We then examine the
reconstructions from eigenpairs by eye with insight from the distance
matrices and regroup PCs whose reconstructions give similar time
series.  We have applied the \emph{k-means} algorithm
\citep{Lloyd:1982} to the set of normalised trajectory matrices
\(\{\mathcal{T}^k\}\) as an aid to grouping.  This is fast and
provides an automated first pass at grouping and the resulting closely
corresponds to the visual impression of the distance matrices.  For
work in this paper, we will present distance matrices by rendering the
elements \(d_{\mu,\nu}\) rather than \(w_{\mu,\nu}\) for
interpretative consistency with our k-means classifier.
\citep{Kalantari.Hassani:2019} advocated using hierarchical clustering
methods which sensible but we have not explored this here. We will
refer to \emph{groups} defined by the distance between reconstructed
trajectories for the eigenpairs.  A number of authors have suggested
Monte-Carlo hypothesis tests \citep{Groth.Ghil:2015} to provide
objective criteria for the statistical significance of the oscillatory
behaviour.  This and other hypothesis testing will be explored as part
of future work.

\subsection{Some simple examples in depth}
\label{sec:sinusoid}

Consider a pedagogical data stream that is six cycles of the cosine
function sampled at 100 evenly-spaced points. We then run the SSA
machinery with window length \(L=15\) to obtain PCs,
eigenvalues, and reconstructions. Figure~\ref{fig:cosineL15} shows the
eigenvalues, PCs, and reconstructions for the pure
sinusoid. Annotations report the normalised cumulative eigenvalues for
the SSA analysis.  As expected more than 97\% of the power is the
first two eigenpairs which represent the oscillation.  The remaining
eigenpairs are insignificant in the reconstruction.  As a case in
point, panels b-e show the first six PCs (paired into three groups).
The first two represent the oscillation and the remaining PCs conspire
to represent the truncation of the signal at the edge of the time
series (e.g. PC3-PC8).  Panel f shows the reconstruction of the time
series for the first group (two PCs) and a comparison of the sum of PC1 and
PC2 with the data. The sum of the reconstructions for PC1 and PC2
reproduces the data.

Now consider a time series that combines a low-frequency cosine with
two cycles of the cosine function combined with a chirp function whose
period increases linearly from three cycles at the beginning of the
sample to none cycles at the end of the interval.  As before, we
create 100 evenly-spaced samples from the
series. Figure~\ref{fig:chirpL15} shows the data, PCs, and
reconstruction for the \(L=15\). In this case, the short window length
mixes the signals together (e.g. PCs 1 and 2 do not form a clean
phased pair that resembles either of the input signals).

The SSA algorithm will attempt to create a series of orthogonal
functions that span the variation in a window length.  For example, if
we consider a series sampled from a simple rectangular (``top-hat'')
function of some width \(s<L\), the algorithm will construct
orthogonal functions in the \(L\)-length interval whose lowest-order
function is a triangle function, followed by antisymmetric, sine-like
function and so on, each orthogonal to the next, in order to reproduce
the input function.  If the rectangle function has length \(s=1\), the
lag-covariance matrix will be the identity, and \(L\) equal weight
orthogonal functions will each be required to reproduce the series.
So, the full scale of the temporal variation must be inside the window
scale to be reconstructed.  In addition, if the function is close to a
continuous orthogonal function in the window, only fewer significant
PCs will be needed for its representation.  In this way, SSA has many
analogous features to Fourier spectral analysis with the added feature
of being adaptive to the time series itself.

Figure~\ref{fig:windows} explores the effect of window length \(L\) on
the first two PCs in the sinusoid and chirp examples. For the sinusoid
example, we only show the first two PCs because the remaining PCs do
not represent signal.  Each period is sampled 16.7 points on average.
We use window lengths that range from 10 to 40 samples, which
illustrates the utility of M-SSA in cross-correlation of the lagged
data.  In all cases the fundamental frequency is well determined.  For
large window lengths, the PCs are distorted as the window hits the
upper boundary of the time series.  For example, the last several
oscillations for \(L=20, 40\) are diminished in amplitude.
Conversely, the \(L=10\) does the best.  Heuristically, \(L\) needs to
be large enough that a signal of interest varies over the window.

For the chirp case, a large window length can nicely separate the
low-frequency from the high-frequency chirp, but the small \(L\)
values mix the signals together. As \(L\) increases, the first two PCs
approach the input cosine function and the remaining PCs represent the
chirp. If the window length is too small to allow the lagged data
replicates in the trajectory matrix to correlate over time, nothing
will be learned from SSA.  Indeed, for a lag of one, the covariance
matrix is simply the variance in the signal and there is one PC: the
time series itself.  Similarly, if the window length is larger than
half of the data length, the correlations at large lag will use very
little of data for cross correlation and will yield nothing of
significance.  Therefore, the window length should be set to some
value larger than the anticipated variation scale of the physical
process and smaller than approximately half of the data length.  Also
note: a large window length leads to a large rank covariance matrix
and large computational expense.  So \(L\to N/2\) should only be used
if necessary.

Figure~\ref{fig:wcorr} is a graphical representation of the distance
matrix defined by equation (\ref{eq:dist}) that helps decide on PC
grouping.  Each element of the matrix indicates the degree of
correlation between the reconstructed time series for each eigenpair.
The maximum distance of 1 obtains for two orthogonal trajectories and
the minimum distance of 0 obtains for two linearly dependent
trajectories.  Orthogonality or lack of any correlation is coded white
and perfect correlation or linear dependency is coded as black.  The
most significant components will be in the upper left and the least
significant components will be in the lower right.  By examining the
normalised sum of the eigenvalues or cumulative variance, we can
restrict our attention to the first \(K^\prime\) rows and columns of
the correlation matrix corresponding to our desired significance
level.  Typically, this matrix divides itself up into blocks of
correlated PCs.  In some unusual cases, the rows and columns make need
to be reordered to recover the natural blocking.  By construction, the
diagonal elements have distance 0.  For our example, the first six
eigenvalues represent 0.939 of the total variance, so we only consider
the \(6\times6\) upper-left block.  This block divides up cleanly into
a \(2\times2\) that represents the cosine signal and a \(4\times4\)
block that represents the chirp (cf. Figure~\ref{fig:chirpL15}).

For EOFs from section \ref{sec:BFE} describing an \(N\)-body
simulation, reconstruction leads directly to density, potential and
force diagnostics. We can use the \(\tilde{a}_{m,i}^{I_j}\) to
reconstruct the underlying potential or density fields in physical
space using equations (\ref{eq:expden}) or (\ref{eq:exppot}).  We will
provide examples in section \ref{sec:barfid}.

\section{Multichannel SSA (M-SSA)}
\label{sec:multi}

\subsection{SSA practice for multiple channels}
\label{sec:mssa_theory}

We can now generalise the SSA to \(M\) time series, here assume to be
\(M\) particular coefficients from equation (\ref{eq:expcof}): the set
\(\mathcal{M} = \{j_1, \ldots\, j_M\}\).  Following the previous
section, denote each time series for the coefficient \(a_j\) as:
\begin{equation}
A_{ji}=(a_{j,i},\ldots,a_{j,i+L-1})^\top, \quad i=1\ldots,K.
\label{eq:mssac1}
\end{equation}
where
\begin{equation}
\mathbf{A}_j = [A_{j1} : A_{j2} : \ldots : A_{jK}].
\label{eq:mssac2}
\end{equation}
Then, the multichannel trajectory matrix \(\mathbb{T}\) may be defined
as
\begin{equation}
\mathbb{T}_M = \left[\mathbf{A}_1, \mathbf{A}_2, \ldots,
\mathbf{A}_M\right].
\label{eq:mssac3}
\end{equation}
The multichannel trajectory matrix has \(KL\) columns with length \(K
= N - L - 1\) (rows).  The covariance matrix of this multichannel
trajectory matrix is
\begin{equation}\small
  \label{eq:traj_d}
  \mathbf{C}_M = \frac{1}{K} \mathbb{T}_M^\top\cdot\mathbb{T}_M
  = \left(
  \begin{array}{lllll}
    \mathbf{C}_{1,1} & \mathbf{C}_{1,2} & \mathbf{C}_{1,3} &\ldots& \mathbf{C}_{1,M}\cr
    \mathbf{C}_{2,1} & \mathbf{C}_{2,2} & \mathbf{C}_{2,3} &\ldots& \mathbf{C}_{2,M}\cr
    \mathbf{C}_{3,1} & \mathbf{C}_{3,2} & \mathbf{C}_{3,3} &\ldots& \mathbf{C}_{3,M}\cr
    \vdots&\vdots&\vdots&\ddots&\vdots\cr
    \mathbf{C}_{M,1} & \mathbf{C}_{L,2} & \mathbf{C}_{L,3} &\ldots& \mathbf{C}_{M,M}\cr
  \end{array}
  \right)
\end{equation}
where each submatrix is
\begin{equation}
  \mathbf{C}_{j,k} =
  \frac{1}{K}\mathbf{A}_j^\top\cdot\mathbf{A}_k.
\end{equation}
Each submatrix \(\mathbf{C}_{j,k}\) has dimension \(K\times K\) as in
the one-dimensional SSA case.  The multichannel covariance both
auto-correlates individual series (as in SSA) and cross-correlates
different time series simultaneously.  Thus, the significance of
temporal information that is common to multiple time series is
reinforced.

The SVD step is the same as in the one-dimensional SSA.  However, each
eigenvector now has a block of length \(K\), one block for each time
series.  Let us denote this as
\[
\mathbf{E}^k = \left[\mathbf{E}^k_1 : \mathbf{E}^k_2 : \ldots :
  \mathbf{E}^k_M\right].
\]
As for standard SSA, we obtain the PCs by projecting the trajectory
matrix onto the new basis as follows:
\begin{equation}
  P^k_i = \sum_{m=1}^M \sum_{j=1}^L a_{m,i+j-1} E^k_{m, j}.
\end{equation}
The PCs are single orthogonal time series that
represent a mixture of all the contributions from the original time
series. It is again useful to define the normalised cumulative sum
of the sorted \(MK\) eigenvalues, where we will use the same notation
as in equation (\ref{eq:eigensum}), but with \(M\) times more eigenvalues,
\begin{equation}
\mathcal{S}_{K^\prime} \equiv \frac{\sum_{k=1}^{K^\prime}
  \lambda_k}{\sum_{k=1}^{MK} \lambda_k},
\label{eq:eigensummssa}
\end{equation} where \(K^\prime<MK\), to quantify the relative
  importance of eigenvectors across channels.

Finally, the last step of the process reconstructs the original time
series of index \(m\in[1,\ldots, M]\) from the PCs as
follows:
\begin{equation}\small
  \label{eq:reconelement}
  \tilde{a}^k_{m,j} =
  \begin{cases} \displaystyle
    \frac{1}{j} \sum_{n=1}^{j} P^k_{n-j+1} E^k_{m,n} & \mbox{if}\ 1\le j < L-1, \\
    \displaystyle
    \frac{1}{L} \sum_{n=1}^{L} P^k_{n-j+1} E^k_{m,n} & \mbox{if}\ L\le j \le N - L + 1 \, \\
    \displaystyle
    \frac{1}{N-j+1} \sum_{n=1-N+M}^{N} P^k_{n-j+1} E^k_m & \mbox{if}\ N-L+2\le j \le N. \\
 \end{cases}
\end{equation}
If we sum up all of the individual PCs, no information is lost; that
is:
\begin{equation}
  \label{eq:reconall}
  a_{m,i} \rightarrow \sum_{k=1}^d \tilde{a}^k_{m,i}
\end{equation}
As described in section \ref{sec:separ}, we often want to sum up the
reconstructions for specific groupings:
\begin{equation}
  \label{eq:recongroup}
  \tilde{a}_{m,i}^{I_j} = \sum_{k\in I_j} \tilde{a}^k_{m,i}
\end{equation}
which gives us the parts of of each coefficient \(a_l(t)\) that
correspond to each dynamical component \(I_j\).

Overall, the M-SSA steps exactly parallel the SSA steps but with more
bookkeeping.  M-SSA rather than the simpler SSA is necessary for our
application to coefficient time series of biorthogonal EOFs.  For any
particular application in galaxy dynamics, the evolution gravitational
field will correlate several basis functions.  So one would have
incomplete signal if one performed \(M\) individual SSAs rather than a
single M-SSA.  However, the more time series we use, the larger the
risk of mixing power between components owing to noise or unsteady
behaviour.  In several of our examples below, we use both EOFs from
the gravitational field and functional representations of the velocity
fields. This mixing is most like amplified when series with different
units are combined.  To combat this, we \emph{detrend} each time
series by subtracting the mean and normalise the time series by the
variance\footnote{Some practitioners advocate more detailed
  detrending, such as polynomial fitting.}.  When combining multiple
types of data, one may weight one type relative to the other.  We have
not done this in our work in section \ref{sec:examples} but it remains
an option.

We may continue to group eigenpairs using the
\(\mathbf{w}\)-correlation or the using k-means clustering and the
Frobenius norm.  For the applications targeted in this paper, the
multiple channels represent spatial features and kinematic signals
induced gravitationally by the same mass distribution, so they will be
naturally coupled.  To aid in grouping, we apply the methods from
section \ref{sec:separ} to each reconstructed channel separately and
look for similarities in the block or group structure as a whole.
Also, we may consider the \(\mathbf{w}\)-correlation matrix and
k-means classification using the distance defined from the matrix norm
of the reconstructed the multichannel trajectory matrix
\(\tilde{\mathbb{T}}\) for each elementary matrix:
\begin{equation}
\tilde{\mathbb{T}}_M^k = \left[\mathcal{T}_1^k, \mathcal{T}_2^k, \ldots,
  \mathcal{T}_M^k\right].
\label{eq:multitraj}
\end{equation}
where each of the \(\mathcal{T}^k_j\) are reconstructed series for the
\(j^{th}\) channel and the \(k^{th}\) eigenpair and
\begin{equation}
  ||\tilde{\mathbb{T}}_M^k||_F =
  \sqrt{
    \sum_{l=1}^M \sum_{i=1}^n \sum_{j=1}^m (\mathcal{T}^k_{lij})^2.
  }
\end{equation}
This is the naturally generalisation of the development in section
\ref{sec:separ}.  Although we have not seen this described elsewhere,
is seems like a natural extension of the \(\mathbf{w}\)-correlation
concept and is useful in practice.  Geometrically, this extended
metric is equivalent to considering vectors the space defined by the
concatenation of time series for all channels.  So this
\emph{multichannel} \(\mathbf{w}\)-correlation describes total
distance between the reconstructed series for each PC for all channels
together.  If all channels mutually take part in a signal of interest,
then combined distance will help discriminate groups by providing more
evidence than any channel individually.  Conversely, independent
channels will form disjoint subspaces and therefore not affect the
distance between group members.  Noisy channels, however, could
degrade separation of groups, so it is up to the practitioner to
choose channels wisely.  Also, it would be possible to weight the
contribution to the total distance by channel based on some prior
assignment of channel importance, but this has not been explored here.

\begin{figure*}
    \centering
    \includegraphics[width=\textwidth]{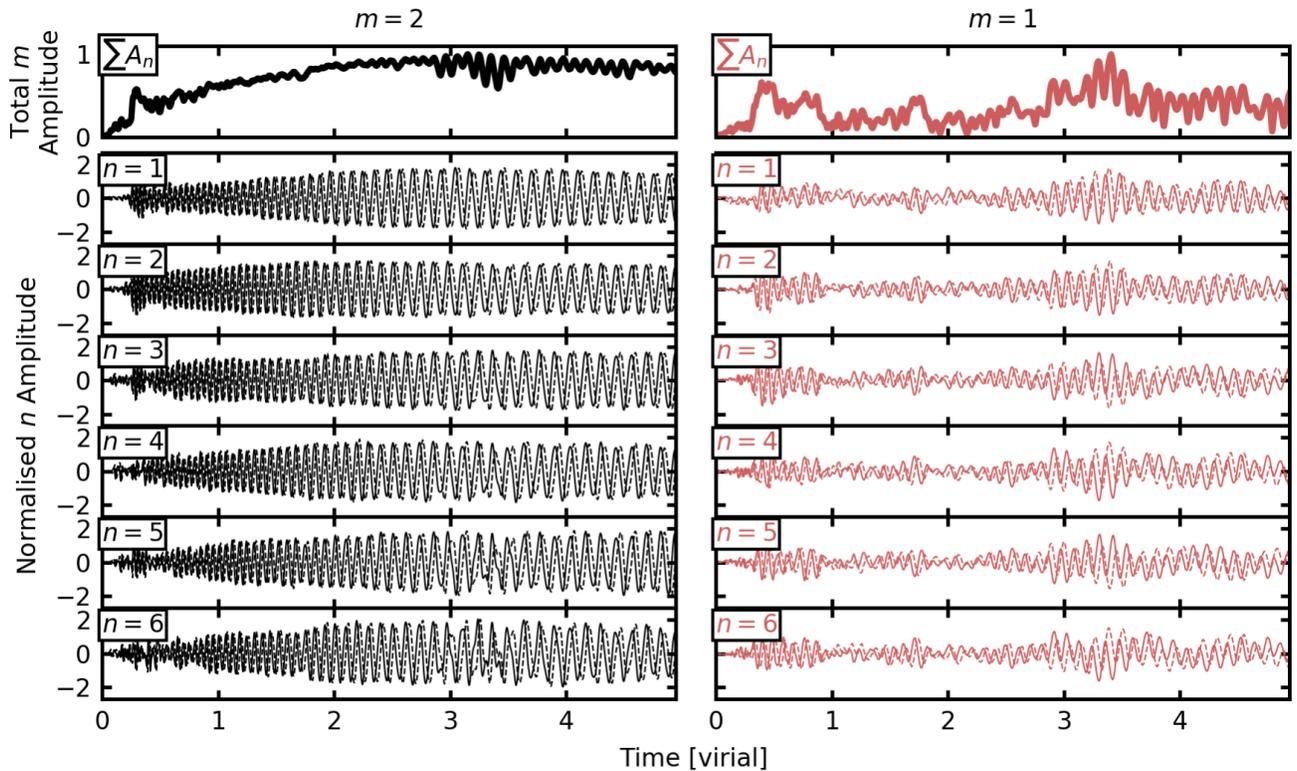}
    \caption{Coefficients of the basis function expansions for
      harmonic orders \(m=2\) (right) and \(m=1\) (left) of the
      fiducial bar simulation. The upper row shows the total modulus
      for the harmonic orders. The remaining panels show the first
      \(n=6\) radial orders in cosine (solid) and sine (dashed)
      series.
      \label{fig:bardata}}
\end{figure*}

\subsection{Other applications of M-SSA}
\label{sec:mssa_tappl}

While our main goal in this paper is the discovery of dynamical
mechanism, M-SSA has a number of other applications to simulations.
The most obvious one is data compression.  In many cases, a small
number of eigenpairs relative to the total number \(MK\) have the
lion's share of the variance; that is: \(S_{K^\prime}\approx 1\) for
\(K^\prime\ll MK\) in the notation of
equation~(\ref{eq:eigensummssa}).  Therefore, we can reconstruct most
of the dynamics with a small number of eigenpairs.  This reduces the
important information in a simulation to very small manageable
summary.  For example, the dominant PCs or their reconstructions might
provide a framework for classifying the dynamical patterns in a suite
of simulations using some distance metric (e.g. the Frobenius norm).

Similarly, we can consider diagnostics of per channel contributions.
Although, we have not seen M-SSA used this way, equations
(\ref{eq:reconelement}) and (\ref{eq:recongroup}) lend themselves to
an estimate of the fraction of each coefficient in any particular
eigenpair or group.  Specifically, let us measure the contribution of
the \(k\)th eigenpair to the \(j\)th coefficient by:
\begin{equation}
  f^k_j \equiv \frac{||\tilde{\mathbf{a}}^k_j||_F}{||\tilde{\mathbf{a}}_j||_F},
\end{equation}
where the Frobenius norm \(||\cdot||_F\) described in
section~\ref{sec:separ}.  By definition \(0<f^k_j<1\) and \(\sum_k
f^k_j=1\). Thus, \(f^k_j\) tells us the fraction of time series \(j\)
that is in PC \(k\).  This tells us whether the particular EOF \(j\)
plays a role in generating the dynamical mechanism. Alternatively, we
may compute
\begin{equation}
  g^k_j \equiv
  \frac{||\tilde{\mathbf{a}}^k_j||_F}{\sum_{j=1}^M|\tilde{|\mathbf{a}}^k_j||_F},
\end{equation}
which is the fraction of PC in series \(j\).  Thus, the histogram
\(g^k_j\) reflects the partitioning of power in the PC \(k\) into the
input coefficient channels \(j\).  So, we can think of this
representation as a single matrix, normed on rows in the case of \(f\)
and normed on columns in the case of \(g\).  In both cases, the norm
over the time series may be restricted to some window smaller than the
total time series.  We have not used this in section
\ref{sec:examples}, but just like compression, it may be useful in the
future for culling degrees of freedom that are informative from those
that are noise-dominated.

\begin{figure}
    \centering
    \includegraphics[width=0.5\textwidth]{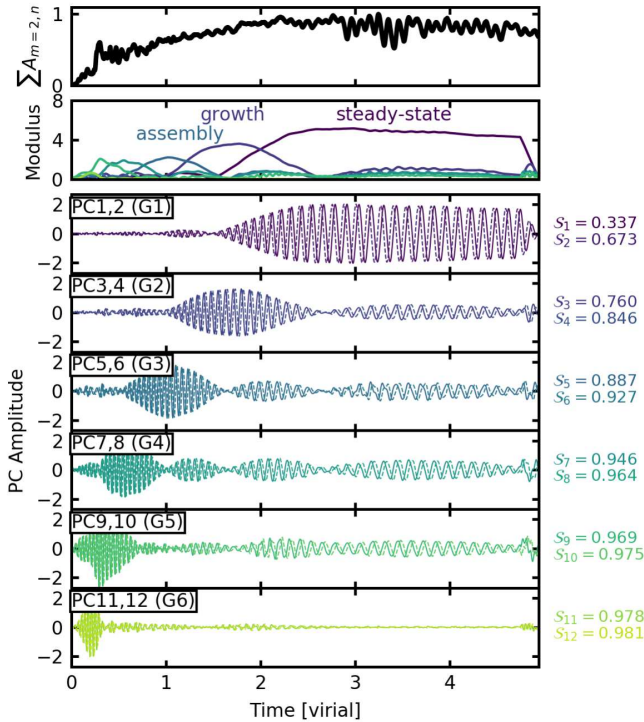}
    \caption{Illustration of M-SSA decomposing the bar simulation into
      physically-informative PCs. The upper row shows
      the sum of the \(m=2\) moduli across all \(n\) orders. The
      second row shows the modulus of each eigenpair of PCs. The phases of bar evolution from
      \protect{\citep{Petersen.etal:2019a}} are labelled. The
      remaining panels show the PC pairs, with cosine-like terms shown
      as solid lines and sine-like terms shown as dashed lines.
      \label{fig:barpcs}}
\end{figure}

\section{Simulation examples}
\label{sec:examples}

We feature the fiducial model from \citet[][hereafter Paper 1, Paper
  2, and Paper3, respectively]
{Petersen.etal:2019a,Petersen.etal:2019b,Petersen.etal:2019c}.  This
model was chosen to be representative of a current-time Milky-Way-like
galaxy.  These papers provided a detailed dynamical analysis of the
possibly interacting channels of secular evolution using orbit
analysis, torques, and harmonic analysis.  Specifically, we found
evidence for three phases in evolution with unique harmonic
signatures.  We recovered known results, such as bar slowdown owing to
angular momentum transfer, and identified a new steady-state
equilibrium configuration and harmonic interaction resulting in
harmonic mode locking.

Because some of these dynamical features will be familiar to many
dynamicists, they provide a pedagogical benchmark for M-SSA.  In the
first example (section \ref{sec:barfid}), we will show that the
simplest application of M-SSA immediately identifies the bar, its
secular growth, and its subsequent pattern-speed decay as a function
of time.  M-SSA effectively compresses the dynamical simulation by
many orders of magnitude into a low-dimensional summary of the key
dynamical feature: the evolution of the bar as a function of time.  In
the second example (section \ref{sec:m12}), we apply M-SSA to the both
the dipole and quadrupole harmonic disc expansion coefficients
simultaneously.  In addition to the \(m=1\) bar coupling reported in
\citet{Petersen.etal:2019b}, M-SSA reveals that the bar drives power
into the natural \(m=1\) mode of the galaxy described in
\citet{Weinberg:94c}.  Finally in sections \ref{sec:kine} and
\ref{sec:buckle}, we explore an extension of basis-function analysis
that combined both our biorthogonal basis functions and a cylindrical
expansion of the velocity fields. In the first velocity example, we
recover the bar-induced velocities; in the second velocity example, we
quantify the velocity signature during a period of
horizontal-vertical coupling.

\subsection{Simple bar analysis from fiducial simulation}
\label{sec:barfid}

This section is a direct application of the development in section
\ref{sec:multi} of our three-dimensional cylindrical disc expansion
described in section \ref{sec:BFE} (see the appendix of Paper 1 for
more detail).  For a quick review, Paper 1 presents a pair of
simulations, one with a cuspy halo and one with a cored halo that are
designed to rapidly form bars.  In the examples considered below, we
will study the cuspy halo from Paper 1 and refer to it has the
fiducial simulation.  An exploratory analysis for an already
well-studied simulation makes for a straightforward identification of
dynamical features in our new method.

The bar dominates the dynamics of the fiducial simulation.  The
strongest dynamical feature of the barred galaxy density and potential
is a rotating quadrupole.  Therefore, a study of the temporal and
spatial features in the \(m=2\) harmonic subspace by M-SSA is a good
first test.  Our simulation records the coefficient time series for
steps of \(h=0.002\) (or approximately every 4 million years in Milky
Way units).  We excerpt the first 12 coefficient time series for the
\(m=2\) harmonic of the cylindrical basis described in section
\ref{sec:BFE}.  These become the \(\hat{a}_j, j=0,\ldots\,11\), the
particle-estimated versions of \(a_j\) in equations
(\ref{eq:mssac1})--(\ref{eq:traj_d}). The run of the \(m=2\)
coefficients is shown in the left panels of Figure~\ref{fig:bardata}.
The slowing bar pattern speed and time-dependent amplitude envelope
are shared by all the series for fixed \(m\). Other distinct dynamical
features are not obvious from this figure.

We now perform the M-SSA on these series.  Computational efficiency
motivates pruning the time series and carefully choosing the window
length, \(L\) in equations (\ref{eq:mssac1}) and (\ref{eq:mssac2}).
First, and especially for the bar-driven dynamics, we may subsample
the time series by striding.  For our case, we use a stride of 2,
every other time step.  Second, we set the window size to \(L=200\);
this implies that we are sensitive to intervals of at most \(\Delta
T=0.4\) time units (or approximately 1 Gyr in Milky Way time units).
We vary \(L\) to check our sensitivity to this choice.  Increasing
\(L\) by a factor of two or four does not substantively change the
dominant PCs.  Our final matrix has rank \(2400\times2400\).  Finally,
the PCs corresponding to the eigenpairs, their eigenvalues, and an
inspection of reconstruction corresponding to each, M-SSA provides an
efficient representation of the dynamics.

For our fiducial bar simulation, approximately 93\% of the total power
is represented in the first six eigenpairs (out of 2400).  Both a by-eye
examination of the distance matrix \(d_{\mu,\nu}\) for the
coefficients that make up the bar using equation (\ref{eq:multitraj})
and the k-means analysis described in sections \ref{sec:separ} and
\ref{sec:mssa_theory} suggests that the first six clusters are pairs
of eigenpairs, which we label as groups 1-6.  As described in sections
\ref{sec:ssaintro}--\ref{sec:multi}, a periodic signal results in
pairs of eigenpairs with similar eigenvalues.  These are analogous to
the sine and cosine terms required to represent a periodic signal in a
Fourier analysis.  SSA does not force any particular PC to have a
constant frequency, but an approximate periodicity will result in
pairs nonetheless.  The dynamics driven by bar formation and evolution
is, of course, driven by the strong quasi-periodic non-axisymmetric
bar potential, so the dominant eigenpairs come in cosine- and
sine-like pairs.  The first
six PC groups are plotted in Figure~\ref{fig:barpcs}.  The lowest-order
group, which dominates the gravitational power, represents the
bar after its formation and growth phases (see
\citealt{Petersen.etal:2019a} for more discussion of bar evolution
\emph{phases}).  The magnitude of any particular pattern identified by
M-SSA will depend on both the strength of the excitation at peak and
the duration of the excitation in the time series.  In this time
series, the simulation spends nearly half of its total time in slow
evolution. Panel b of Figure~\ref{fig:barpcs} combines the cosine- and sine-like components into
a modulus.  This panel succinctly summarises the contribution of each
PC to the bar feature as a function of time.  Also note from Figure~\ref{fig:barpcs} that the bar rotation
period continues to decrease throughout the evolution.  Unlike a
traditional Fourier analysis, M-SSA identifies coherent, persistent
dynamical patterns which may include multiple and evolution
frequencies.

While the first three groups represent 93\% of the total variance
in \(m=2\), there are many other lower-amplitude but significant
features revealed by M-SSA. PC5 + PC6 depict early development of the
\(m=2\) disturbance that will lead to bar formation.  PC7 + PC8, in
particular, depict transient spiral-arm activity.  These are strongest
during bar formation but reoccur at different patterns throughout the
simulation.  The pattern speed of the arms is different than those of
the bar, as expected.

\begin{figure*}
    \centering
    \includegraphics[width=\textwidth]{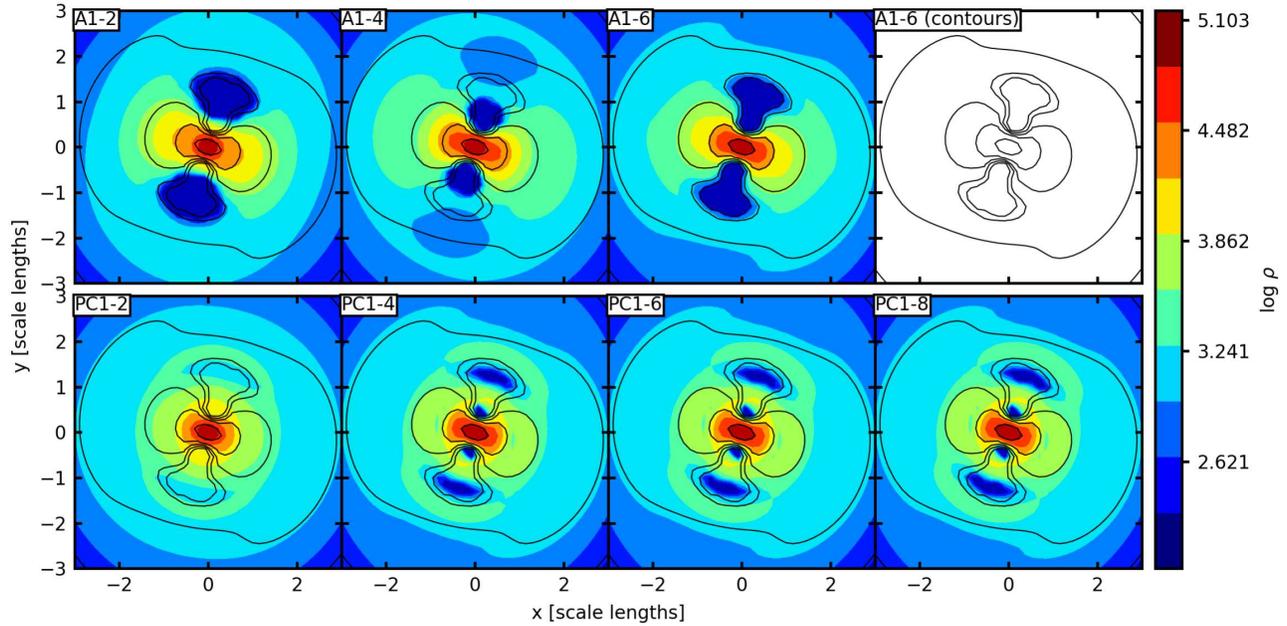}
    \caption{Demonstration of information compression for the \(m=2\)
      example at \(T=2.0\) using M-SSA. Upper: the reconstruction of
      the bar density from the \(m=2\) coefficients (see
      \protect{\ref{fig:bardata}}) as filled contours for radial basis
      functions m through n, denoted as Am--n in each panel. Lower: we
      show the reconstruction of the bar using PCs,
      adding successively more PCs, denoted similarly (see PCs in
      Fig. \protect{\ref{fig:barpcs}}). In all panels, we apply the
      \(m=0\) coefficients from the simulation to approximate total
      density rather than the relative density from only the \(m=2\)
      coefficients. The upper right panel shows only the empty
      contours from the reconstruction of the bar density from
      \(n\le6\), which we then overlay on each other panel for
      comparison. \label{fig:barcompression}}
\end{figure*}

Figure~\ref{fig:barcompression} illustrates the summary nature of
M-SSA by demonstrating a density reconstruction with a small number of
\(m=2\) coefficient series (see Figure~\ref{fig:bardata}) and
time-domain PCs.  We purposely choose \(T=2\) which is during the
transition from the secular growth phase to the steady-state phase.
In addition, we include the \(m=0\) coefficients in the reconstruction
of the bar density to give a better sense of the true density rather
than the density difference. We have investigated a joint \(m=0\) and
\(m=2\) M-SSA computation and find that the \(m=0\) coefficients are
primarily captured in a single PC pair.  This is expected: changes in
the \(m=0\) field are on a secular time scale, which is much longer
than the bar dynamical time.  The upper row in
Figure~\ref{fig:barcompression} shows that the bar is close to
converged after the first four basis functions are coadded.  The
`A1-6' panel is the final bar density reconstruction. The `dimples'
perpendicular to the bar are filled in by higher order harmonics
(principally \(m=4\)) that are not included in this reconstruction.
The lower row in Figure~\ref{fig:barcompression} is the reconstruction
of the bar density from the PCs (see Figure~\ref{fig:barpcs}). From
left-to-right, we add successively more PCs to the
reconstruction. Despite selecting a time when the PCs suggest that the
bar is in a dynamical transition (between growth and steady-state),
the PCs are able to reconstruct the primary features of the bar
density with only two PC pairs. If we chose a time in the centre of a
phase, the bar would be reconstructed with a single PC pair. This
demonstrates the compressive power of M-SSA to reconstruct the bar:
two PC pairs are able to accurately represent the bar, even during a
dynamical transition, while the coefficients require the full
\(n\le6\) radial functions to represent the bar density.

In summary, the M-SSA approach to dynamical analysis of our fiducial
bar simulation in time automatically identified many of the key
results from Paper 1.  Moreover, subdominant but \emph{apparently}
statistically significant features such as spiral arm activity can be
located in time quantified in amplitude.  None of these features could
be inferred from examining Figure~\ref{fig:bardata} alone, although
the information is in these series.  That said, M-SSA does not
\emph{uniquely} identify distinct dynamical components. For example,
the bar itself is spread over multiple PCs in the analysis here.

\begin{figure}
    \centering
    \includegraphics[width=0.5\textwidth]{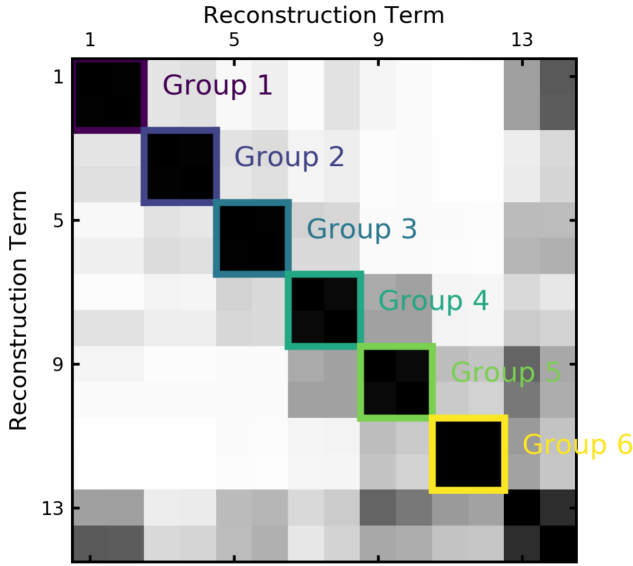}
    \caption{The distance matrix for the \(m=2\) case with low (high)
      distance represented as black (white). The first six groups (all
      pairs of PCs) are labelled.\label{fig:m2wcorr}.}
\end{figure}

\subsection{Combined \(m=1\) and \(m=2\) analysis from fiducial
  simulation} \label{sec:m12}

\begin{figure}
    \centering
    \includegraphics[width=0.5\textwidth]{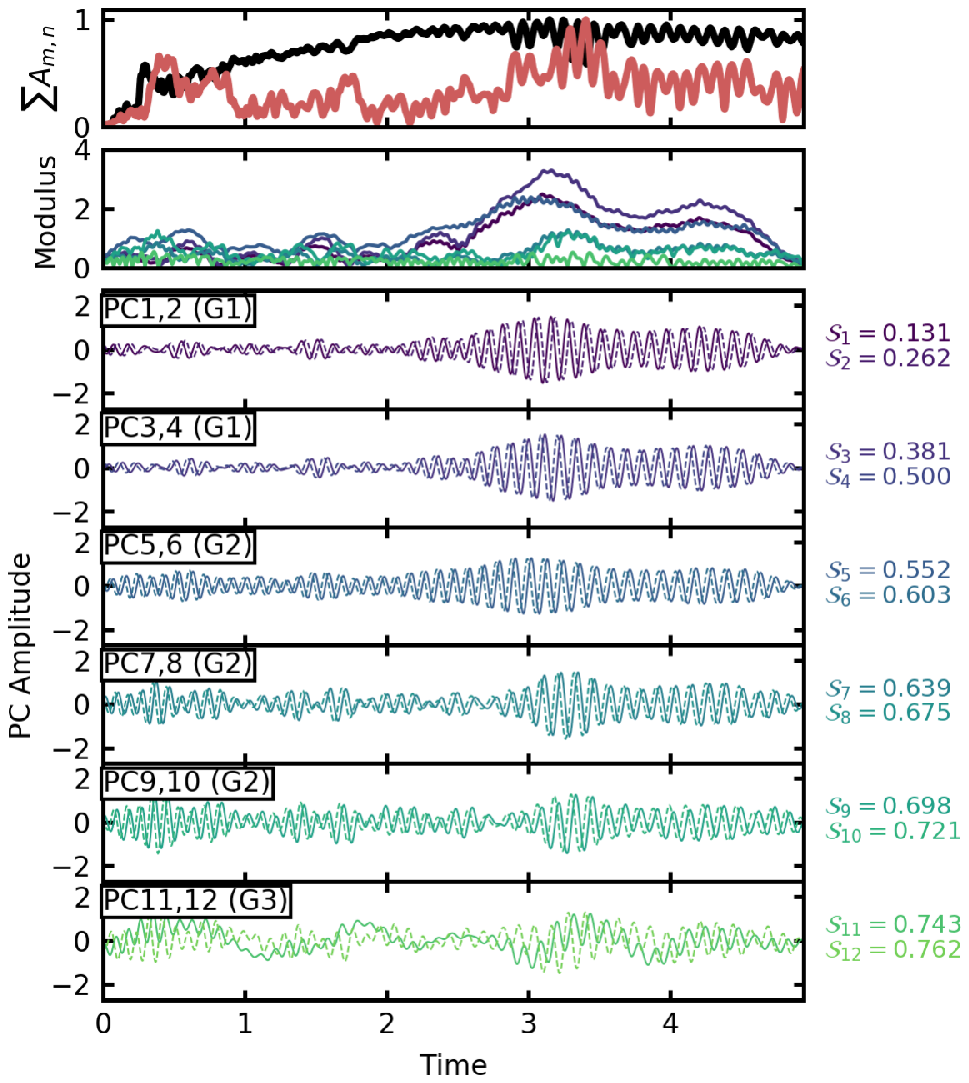}
    \caption{Illustration of M-SSA decomposing the combined \(m=1\)
      and \(m=2\) coefficients from bar simulation into
      physically-informative PCs. The upper row shows
      the sums of the \(m=1\) and \(m=2\) moduli across all \(n\)
      radial orders. The second row shows the modulus of each
      eigenpair of PCs. The remaining panels show the
      PC pairs, with cosine-like terms shown as solid lines and
      sine-like terms shown as dashed lines. The group association for
      each PC pair is listed in the upper left corner of the panel.
      \label{fig:barpcsm1}}
\end{figure}

\begin{figure}
    \centering
    \includegraphics[width=0.5\textwidth]{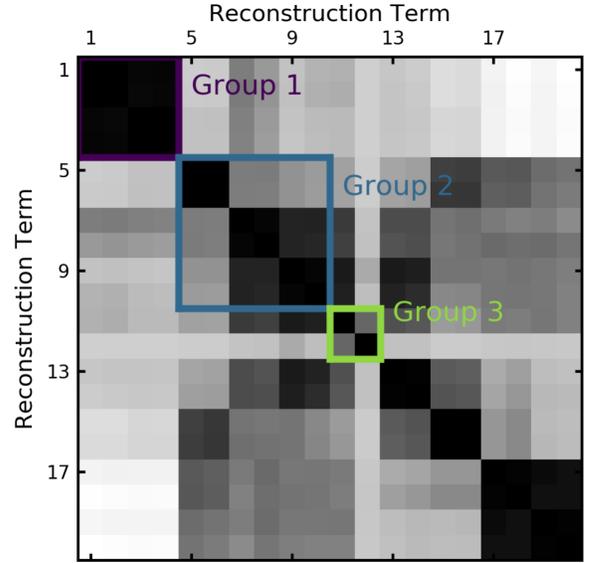}
    \caption{The distance matrix for the \(m=1+m=2\) case with low
      (high) distance represented as black (white). The first three
      groups are outlined and labeled.
      \label{fig:m1m2wcorr}}
\end{figure}

Paper 2 describes an \(m=1\) and \(m=2\) non-linear interaction that
results when the previously independent pattern speeds become
commensurate.  Here, we take a temporally global look at the
interaction using M-SSA.  Specifically, we choose the first six radial
components from each of \(m=1\) and \(m=2\) for \(M=24\) total time
series in equations (\ref{eq:mssac1})--(\ref{eq:traj_d}). The run of
the \(m=1\) coefficients is shown in the right panels of
Figure~\ref{fig:bardata}.  The PCs from the \(m=1+m=2\) analysis are
shown in Figure~\ref{fig:barpcsm1}.  We grouped the individual
eigenpairs by similarity of their PCs and blocks in the distance
matrix (Figure~\ref{fig:m1m2wcorr}; section \ref{sec:separ}).  For all
but one group, these were adjacent pairs of PCs.  One particular group
had three similar eigenpairs.

\begin{figure*}
  \centering \includegraphics[width=\textwidth]{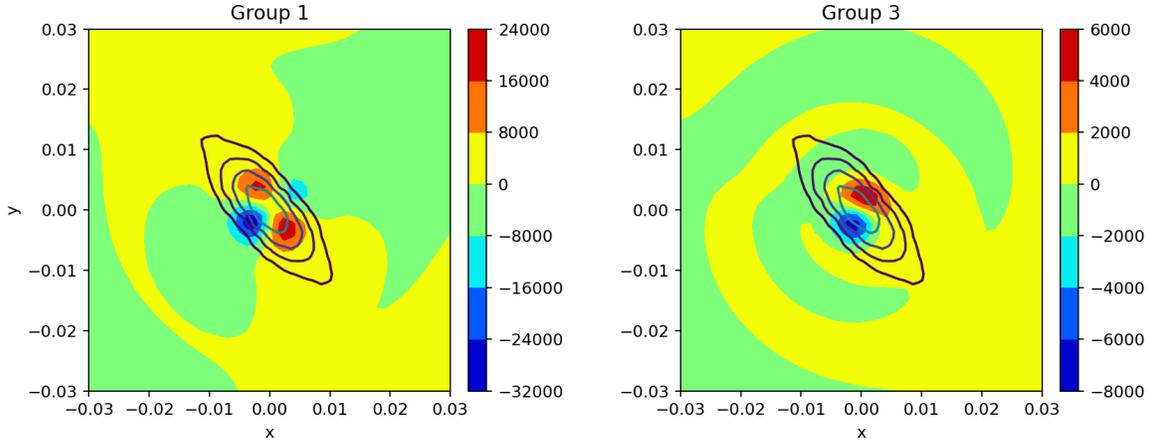}
  \caption{The reconstructed surface density profiles based on the
    combined \(m=1\) and \(m=2\) coefficients for the two eigenpair
    groups described in Figure~\ref{fig:barpcsm1}.  The first group
    shows the quadrupole modulated by a dipole; an animation of
    surface density for the full simulation shows the bar to sloshing
    from side to side.  The left-hand plot shows this as higher
    intensity in direction of negative x, positive y.  The second
    group shows a retrograde \(m=1\) with the very slow oscillation
    frequency (Group 3).  Logarithmically spaced contours in surface
    density is overlaid in the lower row for reference.
      \label{fig:PCm12b}}
\end{figure*}

Overall, the \(m=1\) and \(m=2\) coefficient series are mixed into
multiple PCs. The dominance of both harmonics vary from time to time
in individual PCs but tends to toward a stable split during periods of
similar evolution, i.e. the evolutionary phases described in Paper
1. For example, after the bar forms, PCs1--2 are dominated by \(m=1\)
while PCs3--6 are dominated by bar-like \(m=2\) similar to the previous
subsection.  During the pre-formation, formation and growth phases of
the bar evolution, both \(m=1\) and \(m=2\) coexist.  The density
evolution of each harmonic seems distinct within each phase.  Overall,
PCs1--10 describe a joint \(m=1\) and \(m=2\) modulated interaction with
an coherent pattern speed.  The block consisting of the first four PCs
describe the late-time resonant interaction (Group 1) and the second
block (PC5--10) describes an early-time burst of interaction that
fades after bar growth (Group 2).  As the \(m=1\) and \(m=2\) patterns
become resonant at \(T\approx3\), PC1--4 dominate and the bar appears
to slosh from side to side.  At the same time, PC11--12 appears with
an \(m=1\) signature with a very slow retrograde pattern (Group 3).
This block is significant with respect to the first two blocks with
4\% of the total power. The two-dimensional density profiles for
Groups 1 and 3 are shown in Figure~\ref{fig:PCm12b}.

While the patterns in the primary PCs are clear by eye in a density
animation from the fiducial simulation, detection of the retrograde
pattern required this M-SSA analysis!  We did not see this in the
animation directly.  The existence of weakly damped \(m=1\) modes in
spherical systems was reported in \citet{Weinberg:94c}.  For a
spherical symmetric system, this mode comes in prograde and retrograde
pairs.  Subsequent work showed that in the presence of a disc, the
joint halo and disc mode splits into a very weakly damped retrograde
mode and less weakly damped prograde mode.  So, in this context, the
result found by M-SSA is not a complete surprise. But it was
unanticipated and illustrates the power of M-SSA to find dynamical
correlations that are not evident by visual inspection.

\begin{figure}
  \centering \includegraphics[width=0.5\textwidth]{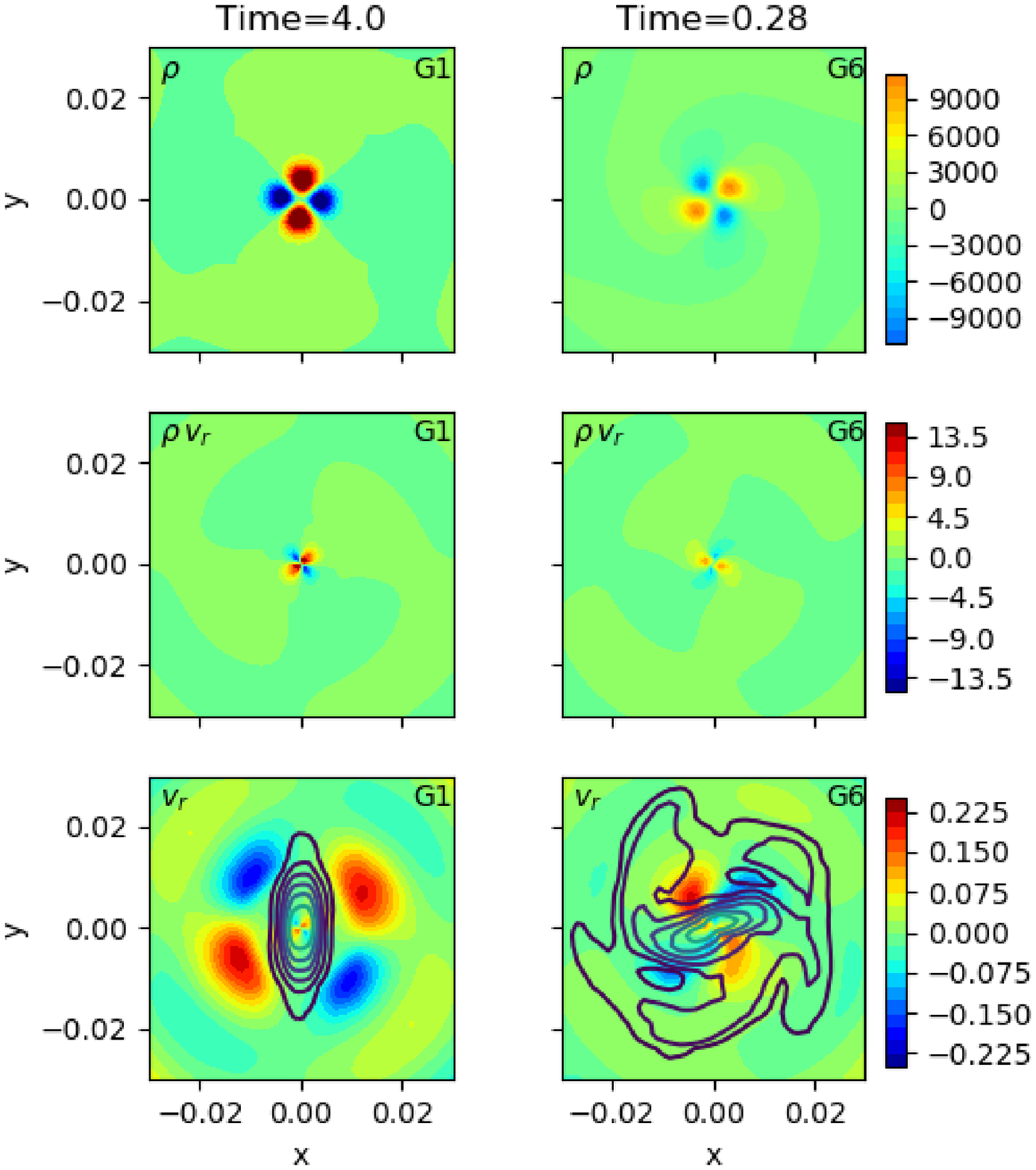}
    \caption{The density (top row), density times velocity (middle row)
      and velocity field (lower row) for two PC groups (columns) in
      the fiducial bar simulation.  Each column shows the PC group
      close to its time of peak amplitude (labeled at the top of each
      column).  The initial scale length of the disc is 0.01 virial
      units and the peak rotation curve is \(\approx1.4\) in virial
      units.  The peak velocity features are 40 km/s when scaled to
      the Milky Way. Logarithmically spaced contours in surface
      density is overlaid in the lower row for reference.
      \label{fig:fid_pc_vr}}
\end{figure}

\subsection{Combined spatial and kinematic signatures from fiducial}
\label{sec:kine}

Although the phase space for any physical system of interest is a
complete description, the history of the density and gravitational
field alone is incomplete; we need the flow or momentum information
for a complete description.  For this reason, the combination of
density and kinematic data together powerfully constrains dynamical
hypotheses while density alone does not.  In addition, density and
kinematics are readily compared with observational data.  Therefore, a
methodology that uses simulations to develop discriminating
statistical tests is desirable, and M-SSA offers this possibility.  As
in sections \ref{sec:barfid}--\ref{sec:m12}, our strategy is a join of
multiple time series of coefficients describing fields of interest.
There are two obvious choices for the join: combining a set of
biorthogonal coefficients from expansions in the form of equations
(\ref{eq:expcof})--(\ref{eq:exppot}) combined with radially local
orthogonal coefficients or radially global orthogonal coefficients
representing the velocity fields.  Locally orthogonal coefficients
include the standard Fourier set at fixed cylindrical radius \(R\),
e.g.
\begin{eqnarray}
c_{mj}(R) &=& \frac{R}{2\pi}\int_0^{2\pi}
d\phi\,\cos(m\phi)\Sigma(R,\phi) v_j, \label{eq:cosv} \\
s_{mj}(R) &=& \frac{R}{2\pi}\int_0^{2\pi}
d\phi\,\cos(m\phi)\Sigma(R,\phi) v_j, \label{eq:sinv}
\end{eqnarray}
where \(v_j\) is some particular component of the velocity in
cylindrical coordinates (\(v_R, v_\phi, v_z\)).  These functions are
orthogonal for all \(m\) at fixed radius \(R\). One might use vector
spherical harmonics \citep[e.g.][]{Hill:1954} to accomplish the same
decomposition for spherically extended systems.  We choose to expand
the product of the surface density and velocity to match the natural
weighting from an N-body simulation.  Evaluated at the positions of
individual particles, \(\Sigma(R, \phi) = \sum_{i=1}^N m_i
\delta(R_i)\delta(\phi_i)/R_i\), equations
(\ref{eq:cosv})--(\ref{eq:sinv}) become:
\begin{eqnarray}
\hat{c}_{mj}(R) &=& \frac{1}{2\pi}\sum_i^N \cos(m\phi_i)m_i
v_{ij}, \label{eq:cosvi} \\
\hat{s}_{mj}(R) &=& \frac{1}{2\pi}\sum_i^N \sin(m\phi_i)m_i v_{ij}.
\label{eq:sinvi}
\end{eqnarray}
We will call this a \emph{ring} expansion for a two-dimensional cylindrical
case and a \emph{shell} expansion for a three-dimensional spherical case.

This approach works, but strong non-axisymmetric disturbances produce
correlated kinematic signatures over a range in radii.  In this
situation, we find that ring and shell expansions are noisy and
inefficient; it spreads the correlated signal over a large number of
series.  An examination of the flow field in a bar immediately
identifies a problem: the \(x_1\) bar orbits move through the
individual rings over very narrow ranges of azimuth.  Therefore the
ring average of \(v_r\) or \(v_t\) is very small.  In this way, rings
dilute the non-axisymmetric flow.  For the fiducial bar model
considered in the sections above, the ring analysis only resulted in
strong signal during times of strong transient disturbance that occur
between evolutionary phases.

To circumvent this bias, we propose using a globally distributed basis
in radius.  The natural choice for a finite radial range is
cylindrical Bessel functions, \(J_n(\alpha_{mj} R)\), for discs and
spherical Bessel functions, \(j_n(\alpha_{lj} R)\), for spherical
distributions, where the constants \(\alpha\) are chosen according the
standard orthogonality conditions \citep{Watson:1966}.  However, it is
better to choose a basis whose lowest-order function resembles the
disc density\footnote{We tested Bessel functions and found that the
  mismatch of the lowest-order function to the disc profile led to
  excessive ringing which clouded the interpretation.}.  To this end
we define a Laguerre basis with functions defined by
\begin{equation}
  G_n(R; a) = \gamma_n(a) e^{-R/a} L_n^{\{1\}}(2R/a)
\end{equation}
where \(g_n=[a^2(n+1)]^{-1/2}\) is the normalisation and
\(L_n^{\{\alpha\}}(x)\) is the generalised Laguerre polynomial
\citep[e.g.][]{Abramowitz.Stegun:64}.  These functions obey the scalar
product:
\begin{equation}
  \left(G_j, G_k\right) = \int dR\,R G_j(R; a) G_k(R; a) = \delta_{jk}.
\end{equation}
Then, the expansion coefficients for any field component
\(v(\mathbf{x})\) may be written as:
\begin{eqnarray}
  C_{mj}[v(\mathbf{x})] &=& n_m \int_0^{\infty} dR R G_j(R;a)
  \int_0^{2\pi} \cos(m\phi) \rho(\mathbf{x}) v(\mathbf{x}), \nonumber \\
  \label{eq:cosBv} \\
  S_{mj}[v(\mathbf{x})] &=& n_m \int_0^{\infty} dR R G_j(R;a)
  \int_0^{2\pi} \sin(m\phi) \rho(\mathbf{x}) v(\mathbf{x}), \nonumber \\
  \label{eq:sinBv}
\end{eqnarray}
where \(n_m = \left[(\delta_{m0} + 1)\pi/2\right]^{-1/2}\) is the
normalisation for trigonometric functions and \(v(\mathbf{x})\) are
one of the cylindrical velocity components \(v_r, v_t, v_z\).
Analogous to the rings, the particle estimators for equations
(\ref{eq:cosBv}) and (\ref{eq:sinBv}) are:
\begin{eqnarray}
  \hat{C}_{mj}[v(\mathbf{x})] &=& n_m \sum_{i=1}^N m_i
  G_j(R_i;a) \cos(m\phi_i) v(\mathbf{x}_i),
  \label{eq:cosBvi} \\
  \hat{S}_{mj}[(v(\mathbf{x})] &=& n_m \sum_{i=1}^N m_i G_j(R_i;a)
  \sin(m\phi_i) v(\mathbf{x}_i).
  \label{eq:sinBvi}
\end{eqnarray}
These coefficients represent the product of the density
\(\rho(\mathbf{x})\) and field quantity \(v(\mathbf{x})\).  We
represent the product \(\rho v\) rather than field alone
\(v(\mathbf{x})\) to facilitate evaluation of the coefficients \(C_{mj},
S_{mj}\) from the particle set. In many cases, we would really like
\(v(\mathbf{x})\) rather than \(\rho v\), and this can be evaluated by
dividing by an independent estimate of \(\rho\) (e.g. a binned
surface density from the particles).

As an example, we used the \(m=2\) time series from section
\ref{sec:barfid} with 8 radial orthogonal functions (\(n\le4\) in
Figure~\ref{fig:bardata}) and the Laguerre expansion with 8 functions.
There are no obvious new features in the PCs as a result of adding the
kinematic time series.  This is confirmed by computing the M-SSA
analysis for the 8 Laguerre coefficient series alone and recovering
nearly the same PCs up to 95\% of the total variance.  However, we do
see that the density and kinematic data together better discriminate
the bar-specific and spiral-arm specific patterns.  We plot the
density, the density times radial velocity, and radial velocity
reconstruction for each PC group near its peak amplitude in
Figure~\ref{fig:fid_pc_vr}.  The velocity fields are computed from the
reconstruction of \(\rho(\mathbf{x})v_r(\mathbf{x})\) divided by the
binned surface density.

The bar is present at the time sampled for groups G1-G5.  Groups G1-G5
are very similar in morphology but peak at different times as
described in section \ref{sec:barfid}; therefore, we only show the
reconstruction from G1 in Figure~\ref{fig:fid_pc_vr}.  The first group
(G1) is the first pair of PCs and represents the bar feature at late
times in its steady-state phase.  Since trapped bar have a prograde
circulation with odd parity relative the bar major axis, we expect the
line of nodes for \(\rho v_r\) to align with the bar density \(\rho\);
this is clearly seen for G1 (Figure~\ref{fig:fid_pc_vr}). The quantity
\(\rho(\mathbf{x})v_r(\mathbf{x})\) depends both on the density of the
disc and its velocity field.  So each of the three quantities
illustrates an aspect of the three interrelated quantities.  For
example, the most obvious kinematic \emph{flow} features follow the
same trend as orbits inside the bar but large radius near the maximum
extent of the trapped bar.  These regions have very low orbit
occupation, however.

The left-hand column in Figure~\ref{fig:fid_pc_vr} shows G6 which
represents a strong \(m=2\) spiral feature before the formation of the
bar; this spiral is prominent in the kinematic pattern.  PCs in this
group represent spiral arm activity at early times
(cf. Figure~\ref{fig:barpcs}), before and during bar formation, and at
later times at lower amplitude with pattern speeds that are lower than
that of the bar.  This suggests that the joint density/potential and
kinematic expansion may be more powerful in some cases than
density/potential expansion alone.  In this case, however, either the
density and kinematic fields determine the PCs on their own.

We have repeated this analysis for \(\rho v_t\) and reproduced the
expected phase-shift in the phase with respect to the density line of
nodes.  This might be expected since \(v_r\) and \(v_t\) are
correlated for the \(x_1\) orbits that dominate the bar structure.
Similarly, the analysis includes \(v_z\).  Our disc EOFs are either
vertically symmetric or antisymmetric with respect to the vertical
midplane (see section \ref{sec:BFE} and \citealt{Petersen.etal:2019a}).
This allows immediate diagnosis of vertical distortions.  To check for
changing vertical structure, we performed the analysis with the
vertically anti-symmetric EOFs together with Laguerre velocity
coefficients: power in anti-symmetric EOFs indicates changing vertical
structure.  Our fiducial simulation does not display any significant
vertical features in the bar and M-SSA recovers this result.

\subsection{Application to vertical coupling}
\label{sec:buckle}

For a disc of constant surface density, the vertical orbital frequency
decreases with increasing scale height.  Depending on the disc and
halo models, the bar pattern speed may approach commensurability with
the vertical orbital frequency as the bar slows. The resulting
resonant coupling exchanges horizontal and vertical action
\citep[e.g.][]{Sridhar.Touma:1996,Quillen:2007} results in an increase of
vertical extent for resonant orbits.  If the resonant orbits are in
phase, the observed signal is a \emph{bobbing} of the orbits above and
below the plane in phase with bar pattern.  While the mechanism
appears to be more closely related to the levitation described by
\citet{Sridhar.Touma:1996}, our goal here is to illustrate the use of
vertical symmetry in both spatial density and velocities to detect and
quantify the vertical feature.  This mechanism called alternatively
called `bending' or `buckling' in the literature
\citep{Raha.etal:1991, Debattista.etal:2006, Sellwood.Gerhard:2020}.
While asymmetries in the bar with respect the disc mid plane produces
a transient vertical asymmetry in some cases, this appears to damp in
time (see \citealt{Petersen.Weinberg:2020}).

\begin{figure}
  \centering 
  \includegraphics[width=0.4\textwidth]{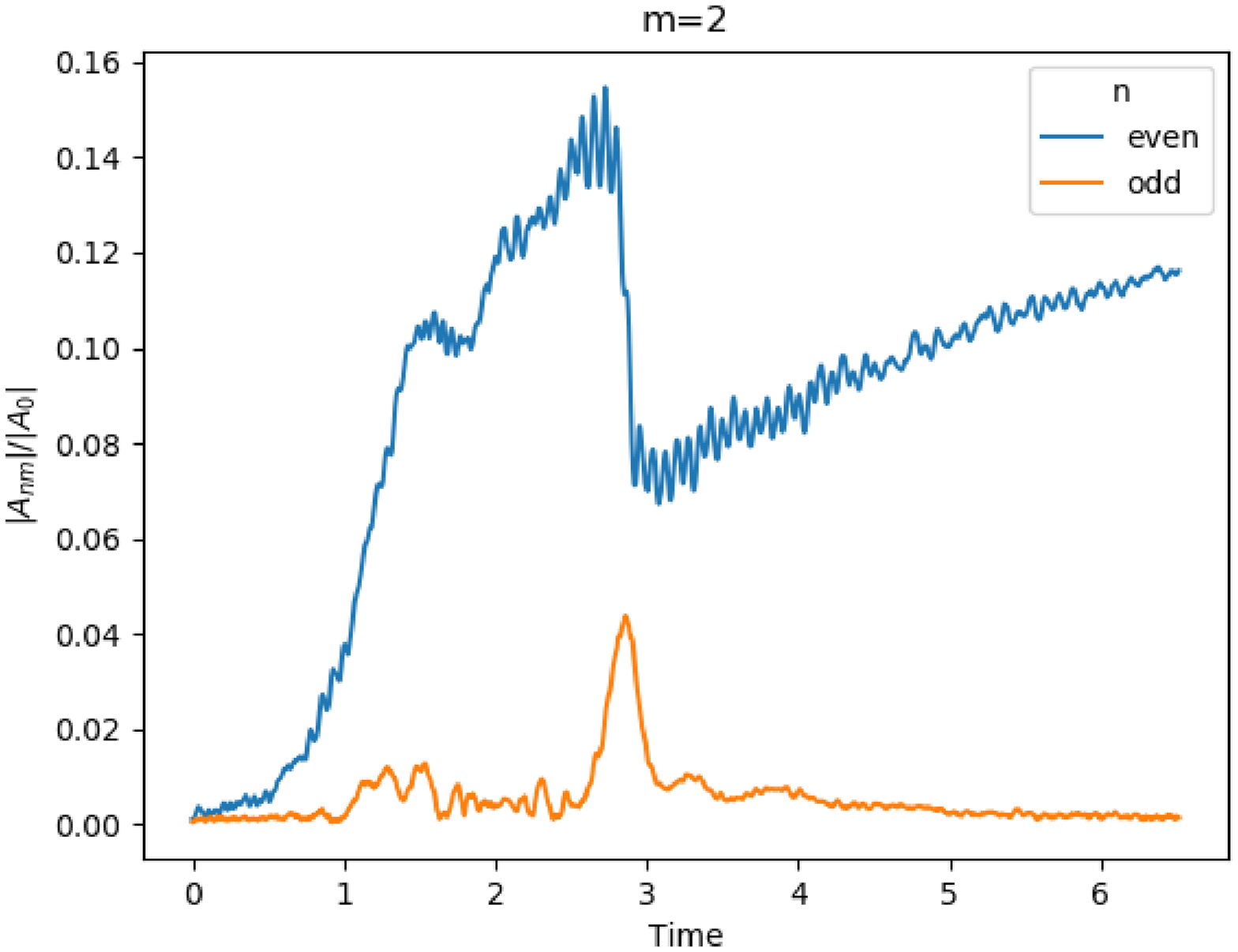}\break
  \includegraphics[width=0.4\textwidth]{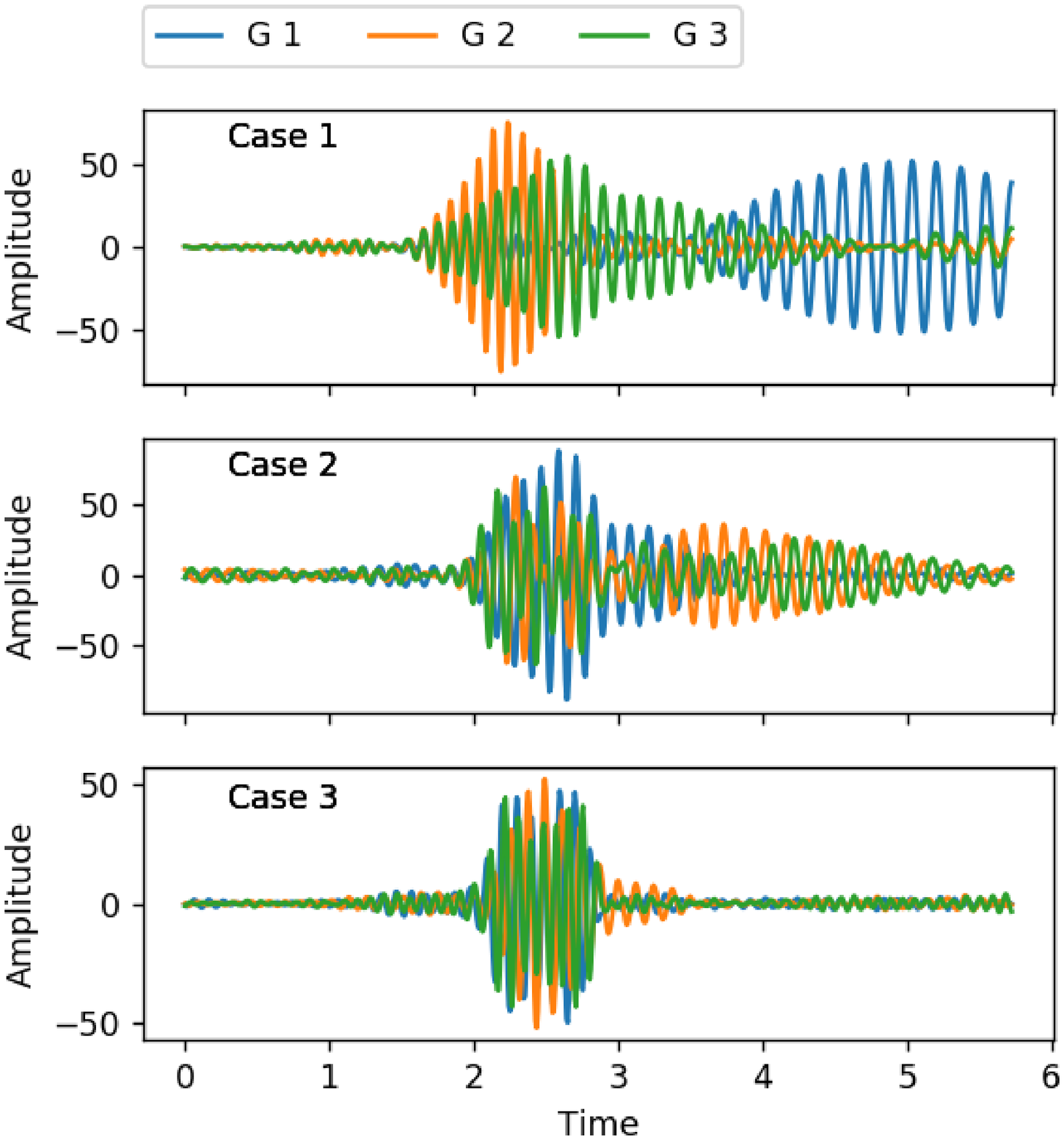}
  \caption{ Top panel: total quadrupole (\(m=2\)) amplitude as a
    function of time for basis functions symmetric (blue) and
    anti-symmetric (orange) about the disc midplane.  The
    vertical-horizontal resonance begins at \(T\approx2\) and
    culminates with jump in anti-symmetric power at \(T\approx2.9\).
    This is followed by a period of increasing bar amplitude that
    correlates with trapping orbits into vertical resonance.  Finally,
    the vertically anti-symmetric asymmetry decays as the orbits lose
    phase coherence.  Remaining panels: the first three PC groups in
    the M-SSA analysis of the first 8 kinematic Laguerre coefficients
    joined with the first 6 vertically symmetric BFE coefficients
    (Case 1), joined with the first 6 vertically anti-symmetric BFE
    coefficients (Case 2) and the Laguerre coefficients alone
    (Case 2).  The first two analyses yield nearly identical
    PCs.  The first two PCs represent 50\% of the total power.
    \label{fig:pow_pc_vz}
  }
\end{figure}

Our example simulation is very similar to the fiducial model discussed
throughout this paper but with a scale height that is two times
larger.  The square root of the quadrupole power or \emph{total
  amplitude} is illustrated in Figure~\ref{fig:pow_pc_vz} (top).  The
horizontal-vertical resonance begins to affect the disk at
\(T\approx2\) with a rapid rise in vertical power beginning at
\(T\approx2.6\) and culminating in a readjustment of the bar support
at \(T\approx2.9\).  Some additional vertical activity persists and
damps away between \(T=4\) and 5.

To describe the vertical evolution using M-SSA, we find the principal
components for the join of the first 8 Laguerre coefficients from
equations (\ref{eq:cosBvi}) and (\ref{eq:sinBvi}) with \(v = v_z\) and
one of (1) the first 6 vertically symmetric BFE coefficients; (2) the
first 6 vertically anti-symmetric coefficients; and (3) no BFE
coefficients.  Let us now compare the results for each of these three
cases.  The first three PC groups for each analysis, representing more
that 50\% of the power in each case are shown in
Figure~\ref{fig:pow_pc_vz} (lower panels).  Recall from
section~\ref{sec:ssaintro} that the PCs represent a mixture of all
time series.

In Case 1, Group 1 represents the long-term evolution the bar after
the vertical evolution without any vertical kinematic correlation.
Group 2 picks out a combination of in-plane changes and vertical
motion associated with the horizontal-vertical resonance that begins
at \(T\approx2\) and rapid grows in amplitude \(T\approx2.6\) to
\(T\approx2.9\).  Group 3 represents the readjustment of the bar that
is correlated with decay of the kinematic vertical power.  Case 2
tells a similar story.  However, the now only the correlations between
anti-symmetric vertical density and kinematics are represented.  The
dominant Group 1 describes the initial coupling between the bar and
vertical motion.  Groups 2 and 3 represent the longer-term decaying
vertical transient. The PCs describing the long-term bar evolution is
absent.  This implies that the vertical coupling has finite duration.
A study of this dynamics suggests that the primary vertical
interaction occurs when the 2:1 azimuthal to vertical frequency
resonance reaches the end of the bar.  This is followed by some
transient trapping.  However, Case 3 differs significantly from the
first two, suggesting that the kinematics alone is insufficient to
uniquely determine the temporal patterns between the spatial and
kinematic variation.  In particular, Case 3 (bottom panel in
Fig.~\ref{fig:pow_pc_vz}) puts emphasis on the vertical motion
resulting from the resonance for all three PC groups.
Higher-order PCs describe the noisy features seen in
Figure~\ref{fig:pow_pc_vz} (top) for \(T\lesssim 1.5\).  The
oscillation in Figure~\ref{fig:pow_pc_vz} (top) corresponds precisely
to the bar pattern; \emph{M-SSA finds no independent pattern speed
  suggesting that there is no separate mode governing bending.}

The differences in the temporal structure revealed by the three M-SSA
case studies illustrate the influence of series selection.  We find
that both vertically symmetric and antisymmetric EOFs will detect the
vertical interaction, but will highlight different facets of the
dynamics.  Case 1 teaches us that in-plane dynamics dominates the
long-term bar evolution.  The analysis reveals lower-amplitude
evidence of the horizontal-vertical coupling at the same frequencies
as the bar, suggesting that the interaction is strongly bar driven.
Case 2 focuses on the vertical structure by correlating the
antisymmetric density and potential excitation with vertical
streaming.  This reveals several stages in the interaction: an initial
resonance which couples the planar angular momentum to vertical
degrees of freedom followed by a slowly decaying transient in vertical
asymmetry.  Case 3 considers vertical velocity streaming only; this
suggests that significant vertical streaming dominates only between
the onset of the resonance but before a structural response.  One must
carefully select series that contain the relevant information for the
dynamical question at hand.

\begin{figure}
  \centering \includegraphics[width=0.35\textwidth]{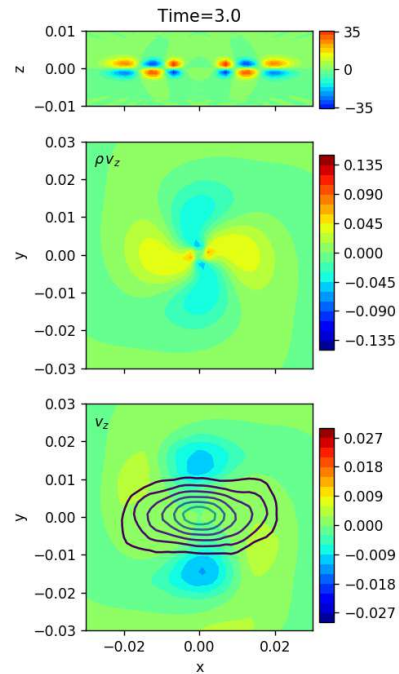}
  \caption{ The density reconstruction \(\rho\) on the \(x\)--\(z\)
    plane (top), \(\rho v_z\) (middle) and velocity \(v_z\) (bottom)
    for PC1 at \(T=3.0\) in Case 2. This reveals an ensemble of bar orbits that
    describe a correlated vertical oscillations as they circulate
    through the bar.  PC2 (not shown) describes the same pattern.
    Logarithmically spaced contours of disc surface density are
    overlaid in the bottom panel for comparison.
    \label{fig:pc_recon1}
  }
\end{figure}

Figure~\ref{fig:pc_recon1} compares the reconstruction of the velocity
signal, \(\rho(\mathbf{x}) v_z(\mathbf{x})\) and \(v_z(\mathbf{x})\),
with the vertically anti-symmetric component of disc density
\(\rho(\mathbf{x})\) from Case 2.  Large positive or large negative
values of \(\rho(\mathbf{x}) v_z(\mathbf{x})\) (middle panel) and
\(v_z(\mathbf{x})\) (bottom panel) indicate regions dominated by
significant upward and downward mean orbital motions, respectively.
Consider a bar-supporting orbit in the frame of the bar rotation.
Figure~\ref{fig:pc_recon1} tells us that a representative orbit moves
upwards as it moves from apocentre to pericentre and downwards from
pericentre to apocentre.  Figuratively, the orbits follow a
saddle-shaped warped surface whose line of nodes aligns with the bar.
Comparing all three panels, we see that the vertical \(m=2\) feature
is characterised by the inner bar moving in the opposite direction
vertically as the outer bar.  The velocity streaming is dominated by
vertical motions in the outer bar.  The top panel shows a cut through
the \(x\)--\(z\) plane.  The multiple sign-changes in the vertical
density arises from the in-phase oscillation of orbits near the
horizontal--vertical resonance at different radii owing to interaction
with the existing transient structure.

M-SSA reveals that the bar is not bending dynamically but stays bent
after the initial vertical coupling, slowly settling back to the disc
mid plane in time.  This pattern is preserved for many bar rotations.
By symmetry, orbits that downwards as they move from apocentre to
pericentre and upwards from pericentre to apocentre should be equally
likely.  This suggests that some transient or phase dependence is
responsible for the sign of the vertical bobbing.  Indeed, the decay
of the anti-symmetric power seems to correspond to a homogenisation of
these two perpendicular phases.  After mixing, the bar has the
characteristic boxy or peanut shape.  We will use these tools to
describe the dynamics of these vertical features in a later paper.

\section{Discussion and summary}
\label{sec:discuss}

This paper proposes a general approach for discovering dynamical
mechanisms in simulations using a combination of basis-function
expansions (BFEs) and multichannel singular spectrum analysis (M-SSA),
with the possibility of additional time series describing associated
field data such as kinematics, age or chemical data.  While our
examples are simulations that use the BFE method for solution of the
Poisson equation, any set of simulation snapshots may be post
processed to generate BFE coefficients.

We applied these ideas several situations that we have already
studied in detail as method tests.  We were pleasantly surprised that
M-SSA not only recovered and reconfirmed our prior results, it
provided new discoveries and insights.  Our use of M-SSA on BFE time
series of coefficients might be considered a form of unsupervised
learning.  We emphasise that no prior information about the spatial or
temporal nature of these dynamics are included in our analysis: the
only tuning parameters are the length of the time series, mostly
determined by the astronomical scenario itself, and the window length,
which is chosen to be less than half of the entire duration and often
smaller for computational expediency.  M-SSA \emph{rediscovered} the
three epochs from \citet{Petersen.etal:2019b} that were determined by
conventional methods.
    
Simulations of disc galaxies reveal natural \(m=1\) or \emph{seiche}
and \(m=2\) frequencies in nearly all cases.  The seiche excitation is
less dramatic than a bar or spiral arms but rather looks like a
central offset or lopsided density feature with a pattern slower than
most orbital frequencies. When applied to a combination of \(m=1\) and
\(m=2\) coefficients simultaneously, M-SSA revealed the same dynamics
discovered by eye: namely that when the bar pattern frequency
approaches the natural dipole frequency, the bar and the seiche
exchange power.  In addition, M-SSA discovered a slow retrograde
oscillation of the disc that is excited by the \(m=1\) and \(m=2\)
interaction.  The pattern speed is very slow, corresponding to a
gigayear period scaled to the Milky Way.  Such modes were predicted by
\citet{Weinberg:94c} but was not obvious in the simulation by eye.
While the power in this slow retrograde mode is only 5\% of the total,
its detection is highly significant.

We show that BFE time series may be augmented by other quantities.
Kinematic data, which are easy to compute from simulations and observe
directly from spectroscopic surveys, are included here using a
Fourier-Laguerre expansion of the cylindrical velocity vector.  We
show that the spatial and kinematic data reinforce each other as we
expect they should given the nature of phase space.  Moreover, the
vertical velocity field combined with anti-symmetric terms of the disc
expansion basis allow clear description of the bar `buckling'
phenomenon.

Although we have not explored additional field data besides kinematic
here, many codes simulate star formation and track chemical and
stellar age data.  These fields may be easily added to investigate the
coupling between chemical and star formation patterns and specific
dynamical mechanisms such as minor mergers, spiral arm or bar
formation.  Other promising time series include conserved quantities
of particular orbit ensembles such as approximate action values or
energies in order to test or elicit dynamical mechanisms.

Our implementation of of M-SSA for these tests is standard and other
publicly available implementations can be found in several published
packages, such as \emph{The Singular Spectrum Analysis - MultiTaper
  Method (SSA-MTM)} distributed by the Department of Atmospherical and
Oceanographic Sciences at UCLA, commercial packages, among others.
Our implementation naturally interfaces with our EXP code
\citep{Petersen.etal:2019a} and integrated tools will be provided as
part of an upcoming public release.

\section*{Data availability statement}
The simulation data underlying this article will be shared on
reasonable request to the corresponding author.  All other
data underlying this article are available in the article.

\section*{Acknowledgements}
MDW thanks the Center for Computational Astrophysics at the Simons
Flatiron Institute for visitor support in 2018 during which this
project was conceived.  This research was also supported in part by
the National Science Foundation under Grant No. NSF PHY-1748958 during
a visit to KITP.  MSP acknowledges funding from the UK Science and
Technology Facilities Council (STFC). We are grateful to Kathryn
Johnston for comments on an early version of this manuscript.

\bibliographystyle{mn2e}

\label{lastpage}

\end{document}